\documentclass[%
 superscriptaddress,
 reprint,
 showpacs,preprintnumbers,
 nofootinbib,
 amsmath,amssymb,
 aps,
 prd,
 floatfix,
]{revtex4-2}

\usepackage{graphicx}

\usepackage{xcolor}
\usepackage{lineno}
\usepackage{adjustbox} 

\usepackage{amssymb,amsmath,amstext,amsthm,amsfonts,amsfonts}
\usepackage[utf8]{inputenc}
\usepackage[linesnumbered,ruled]{algorithm2e}
\usepackage{url}
\usepackage{soul}
\usepackage{bm}
\usepackage[normalem]{ulem}
\usepackage{tabularx}
\usepackage{multirow, makecell}
\usepackage{floatrow}
\usepackage{comment}

\usepackage{color, colortbl}
\definecolor{Gray}{gray}{0.9}
\definecolor{LightCyan}{rgb}{0.88,1,1}
\usepackage[first=0,last=9]{lcg}

\newcommand{\CC}{\cellcolor{white}}

\usepackage{amsmath}

\newfloatcommand{capbtabbox}{table}[][1.275\linewidth]

\newcommand{\review}[1]{\textcolor{black}{#1}}

\newcommand{\affiliationETHZ}{ETH Zurich, Institute for Particle physics and Astrophysics, CH-8093 Zurich, Switzerland.}
\newcommand{\affiliationSBU}{State University of New York at Stony Brook, Department of Physics and Astronomy, Stony Brook, New York, USA.}

\DeclareMathAlphabet{\mathsfit}{\encodingdefault}{\sfdefault}{m}{sl}
\SetMathAlphabet{\mathsfit}{bold}{\encodingdefault}{\sfdefault}{bx}{n}

\begin{document}

	\title{Artificial intelligence for improved fitting of trajectories of elementary particles \\
	in inhomogeneous dense materials immersed in a magnetic field} 

    \author{Sa\'ul Alonso-Monsalve}
    \email[E-mail: ]{saul.alonso.monsalve@cern.ch}
    \author{Davide Sgalaberna}
    \author{Xingyu Zhao}
    \affiliation{\affiliationETHZ}
    
    \author{Clark McGrew}
    \affiliation{\affiliationSBU}
    \author{Andr\'e Rubbia}
    \affiliation{\affiliationETHZ}

\begin{abstract}
\noindent

In this article, we use artificial intelligence algorithms to show how to enhance the resolution of the elementary particle track fitting 
in inhomogeneous dense detectors, such as plastic scintillators. We use deep learning to replace more traditional Bayesian filtering methods, drastically improving the reconstruction of the interacting particle kinematics.
We show that a specific form of neural network, inherited from the field of natural language processing, is very close to the concept of a Bayesian filter that adopts a hyper-informative prior.
Such a paradigm change can influence the design of future particle physics experiments and their data exploitation.

\end{abstract}

\maketitle

%

\section{Introduction}
\label{sec:introduction}

Understanding the behaviour of subatomic particles traversing dense materials, often immersed in magnetic fields, has been crucial to their discovery, detection, identification and reconstruction, 
and it is a critical component for exploiting any particle detector~\cite{HASERT1973121,HASERT19741,201230,20121,PhysRevLett.74.2626,GRUBER2020162025}.
Modern radiation detectors have evolved towards ``imaging detectors'', in which elementary particles leave individual traces called ``tracks''~\cite{AMERIO-icarus,Abi:2020evt,Acciarri:2016smi,Blondel_2018,calice}.
These imaging detectors 
require a ``particle flow'' reconstruction: 
particle signatures are precisely reconstructed in three dimensions, and the kinematics (energy and momentum vector) of the primary particle can be measured track-by-track.
It also means that a more significant amount of details can be obtained on each particle. These features open the 
question of which methods are best suited to handle the ``images'' created by the subatomic particles.

Common Monte Carlo (MC) based methods used in the track fitting flow belong to the family of Bayesian filters and, more specifically, they are extensions to the standard Kalman filter~\cite{kalmanfilter} or particle filters algorithms, with special mention to the Sequential Importance Resampling particle filter (SIR-PF)~\cite{sirpf}.
The knowledge about how an electrically charged subatomic particle propagates through a medium (i.e., the energy loss, the effect of multiple scattering, and the curvature due to magnetic field)
can be embedded into a prior (often in the form of a covariance matrix for Kalman filters). 
In particle filters, the nodes of the track are fitted sequentially: given a node state, the following node in the particle track is obtained by throwing random samples - known as ``particles'' - and making a guess of the following state by applying a likelihood between the sampled particles and the data (which could be, for instance, the signatures obtained from the detector readout channels).
The result can be the position of the fitted nodes of a particle track or directly its momentum vector and its electric charge.
Usually, the problem is simplified using a prior that follows a Gaussian distribution, like in the Kalman filter, which also considers a simplified version of the detector geometry.
Examples can be found in 
\cite{Innocente:1991md,Innocente:1993kz,Cervera-Villanueva:2004dba}. 
%
However, the filtering is not trivial since both the particle energy loss and multiple scattering angles depend on the momentum, which changes fast in dense materials, and approximations are often necessary.
Moreover, it is hard to incorporate finer details of a realistic detector geometry and response (e.g., signal crosstalk between channels, air gaps in the detector active volume, presence of different materials, or non-uniformities in the detector response as a function of the particle position, inhomogeneous magnetic field) or to deal with deviations in the particle trajectory due to the emission of high-energy $\delta$-rays, with photon Bremsstrahlung emission, with the Bragg peak of a stopping particle, or with inelastic interactions.
All these pieces of information are available in the simulation of a particle physics experiment 
\cite{AGOSTINELLI2003250,1610988,ALLISON2016186,Ahdida:2022gjl,Battistoni:2015epi} 
and can be validated or tuned with data 
but it is not straight-forward to use them in the reconstruction of the particle interaction.
Hence, developing new reconstruction methods capable of analysing all the information available becomes essential.

The most promising solution 
is given by artificial intelligence and, more specifically, by deep learning, a sub-field of machine learning based on artificial neural networks~\cite{de2017learning, Radovic-et-al-2018-machine, doi:10.1146/annurev-nucl-101917-021019, RevModPhys.91.045002}. Initially inspired by how the human brain functions, these mathematical algorithms can efficiently extract complex features in a multi-dimensional space after appropriate training. Neural networks (NNs) have been found to be particularly successful in the reconstruction and analysis of particle physics experiments~\cite{baldi2014searching, dunecvn, https://doi.org/10.48550/arxiv.2102.01033, andrews2020end, PhysRevD.103.032005}.
%
%
%
Thus far, deep learning has been used in high-energy physics (HEP) for tasks such as classification~\cite{novacvn,dunecvn,nguyen2019topology,Bhattacharya2020}, 
semantic segmentation~\cite{MicroBooNE-2020-semantic,PhysRevD.103.032005}, 
or regression~\cite{Cheong_2020,QIAN2021165527,carloni2022convolutional}.
Typically, the raw detector signal is analysed
to extract the physics information.
This approach is quite common in experiments studying neutrinos, for example, to classify the flavour of the interaction ($\nu_{\mu}$, $\nu_e$, or $\nu_{\tau}$) by using convolutional neural networks (CNNs) \cite{novacvn,dunecvn,Acciarri:2016ryt},
or the different types of signatures observed in the detector~\cite{MicroBooNE-2020-semantic,PhysRevD.103.032005}.
These methods have been shown to outperform more traditional ones, such as likelihood inference or decision trees. 
However, asking a neural network to extract high-level physics information directly from the raw signatures left in the detector by the charged particles produced by a neutrino interaction is conceivable as challenging. An example is the neutrino flavour identification (as mentioned before), which incorporates diverse contributions, from the modelling of the neutrino interaction cross-section to the propagation of the particles in matter and, finally, the particular response of the detector. Expecting a neural network to learn and parametrise all these contributions could become unrealistic and lead to potential deficiencies.


An alternative and promising approach is to use deep learning to assist the more traditional particle flow methods in reconstructing particle propagation, which consists of a chain of different analysis steps that can include the three-dimensional matching of the 
voxelised signatures in the 
detector readout 2D views,
the definition of more complex objects such as tracks 
and, finally, the fit of the track in order to reconstruct the particle kinematics.
As described above, the last step is critical and is usually performed by a Bayesian filter that has to contain as much information as possible in its multi-dimensional prior.
It becomes clear that, overall, the reconstruction performance depends on the detector design (e.g., granularity or detection efficiency) and on the a priori knowledge of the particle propagation in the detector, the prior.
Although prohibitive for traditional Bayesian filters, the problem of parameterising a high-dimensional space can be overcome with deep learning since neural networks can be explicitly designed for it.

Even though the generic idea of using deep learning as an alternative to Bayesian filtering has already been explored~\cite{GAO2019279}, common applications focus on tasks such as enhancing and predicting vehicle trajectories~\cite{s20185133, 8713418}. Furthermore, the closest application we can currently find in HEP and other fields like biology is to use deep learning to perform ``particle tracking''~\cite{dezoort2021charged, 10.1093/bioinformatics/btaa597, Tsaris_2018}, which relies on connecting detected hits to form and select particles, distinct from the idea of fitting the detected hits to obtain a good approximation to the actual particle trajectory.

In this article, we propose the design of a recurrent neural network (RNN) and a Transformer to fit particle trajectories. We found that these neural nets, inherited from the field of natural language processing, are very close to the concept of a Bayesian filter that adopts a hyper-informative prior. Hence, they become excellent tools for drastically improving the accuracy and resolution of elementary particle trajectories.

\begin{figure*}[htb] 
\centering
\includegraphics[width=1.\textwidth]{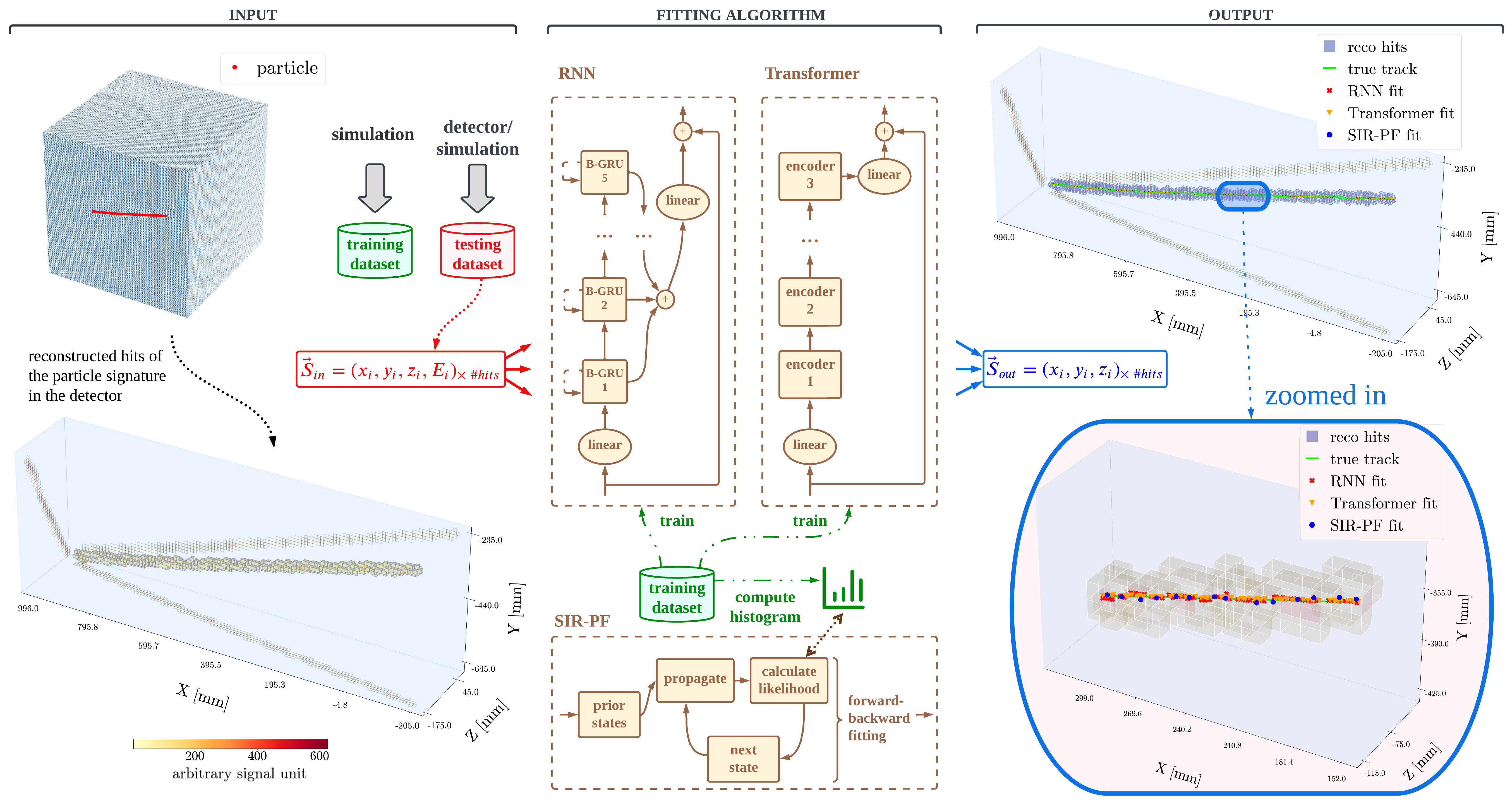}
  \caption{\label{fig:workflow}
  Workflow of a crossing muon track fitting using the three algorithms: recurrent neural network (RNN), Transformer, and Sequential Importance Resampling particle filter (SIR-PF). From left to right, the diagram shows the steps from the particle simulation/detection until the particle is fitted using the different algorithms.}
\end{figure*}

\section{Proof-of-principle}
\label{sec:proof-of-principle}


In order to train and test the developed neural networks and compare their performance with a more classical Bayesian filter, an idealized three-dimensional fine-grain plastic scintillator detector was taken as a case study. We simulated a cubic detector composed of a homogeneous plastic scintillator with a size of $2\times2\times2$ $m^3$. A uniform magnetic field is applied, aligned to one axis of the detector (X-axis) and its strength is chosen to be 0.5~T. The detector is divided into small cubes of size 1 cm$^{3}$, summing $200 \times 200 \times 200$ cubes in total. Each cube is assumed to be equipped with a sensor that collects the scintillation light produced when a particle traverses it. We simulate the signals read from each sensor and reconstruct the event based on these signals. The track input to the fitters will be extracted from event reconstruction.

Overall, the simulation and reconstruction are divided into three steps:
\begin{enumerate}
    \item \textbf{Energy deposition simulation}: this step uses the Geant4 toolkit \cite{AGOSTINELLI2003250,1610988,ALLISON2016186} to simulate particle trajectories in the detector and their energy deposition along the path.
    \item \textbf{Detector response simulation}: this step simulates detector effects and converts the energy deposition into signals the detector can receive. The current detector effect being considered is the light leakage from one cube to the adjacent one (named crosstalk). The leakage probability per face is assumed to be 3\%. The energy deposition is converted from the physics unit (MeV) into the ``signal unit'' (depending on the detector) by using a constant factor, which is fixed to be 100 / MeV for this analysis. Besides, a threshold is also implemented on the sensor, requiring that at least one signal unit be received to activate the sensor.
    \item \textbf{Reconstruction}: this step takes the signals generated from the former steps and reconstructs objects, such as tracks, that can be input to the fitter. Starting from 3D ``cube hits'' (what we have after the detector response simulation), we then apply the following two methods to find track segments from the whole event: (1) the Density-Based Spatial Clustering of Applications with Noise (DBSCAN)~\cite{dbscan}, which groups hits into large clusters that, in each cluster, all hits are adjacent to each other; (2) the minimum spanning tree (MST)~\cite{kruskal1956shortest} for each cluster to order hits and break the cluster into smaller track segments at each junction point. Afterwards, the primary track segment will be selected for track fitter input.
\end{enumerate}


The simulation and reconstruction processes resulted in single-charged particles (protons, pions $\pi^{\pm}$, muons $\mu^{\pm}$, and electrons $e^{\pm}$) starting at random positions in the detector active volume with isotropic directions and uniform distributions of their initial momentum: between 0 and 1.5 GeV/c (protons), 0 and 1.5 GeV/c (pions), 0 and 2.5 GeV/c (muons) and 0 and 3.5 GeV/c (electrons). Each particle consisted of a number of reconstructed 3D hits belonging to the track, where each hit is represented by a three-dimensional spatial position and an energy deposition in an arbitrary signal unit. For each reconstructed hit in a particle, there is a \textbf{true node} (to be learnt during the supervised training) which represents the closest 3D point to the hit in the actual particle trajectory; in that way, there is a 1-to-1 correspondence between reconstructed hits (even for crosstalk) and true nodes. We refer in the rest of the article to the output of the algorithms developed as \textbf{fitted nodes}, which form the fitted trajectory for each particle.


\section{Results}
\label{sec:results}

In this section, we discuss the performance of a recurrent neural network (RNN)~\cite{JORDAN1997471, 10.5555/553011, SHERSTINSKY2020132306} and a Transformer~\cite{Vaswani2017attention}, comparing their results with the ones from a custom SIR-PF (as described in Sec.~\ref{sec:introduction}).
The developed methods, described in detail in Sec.~\ref{sec:methods}, were run on a test dataset of simulated elementary particles
(statistically independent of the dataset used for training) 
consisting of 1,895,877 particles (422,836 protons, 490,003 pions $\pi^{\pm}$, 492,560 $\mu^{\pm}$, and 490,478 $e^{\pm}$). For each simulated particle, the goal was to use the reconstructed hits to predict the actual track trajectory and then to analyse its physics impact on the detector performance, as described later in this section. 
The output of the different methods was a list of fitted nodes, i.e. the predicted 3D positions of the elementary particle in the detector.
A visual example of the particle trajectory fitting using the different techniques is shown in Fig.~\ref{fig:workflow}.

\subsection{Fitting of the particle trajectory}
\label{sec:performance}

\begin{figure*}[htb]
    {\adjincludegraphics[height=6.9cm,trim={{.025\width} 0 {.05\width} 0},clip]{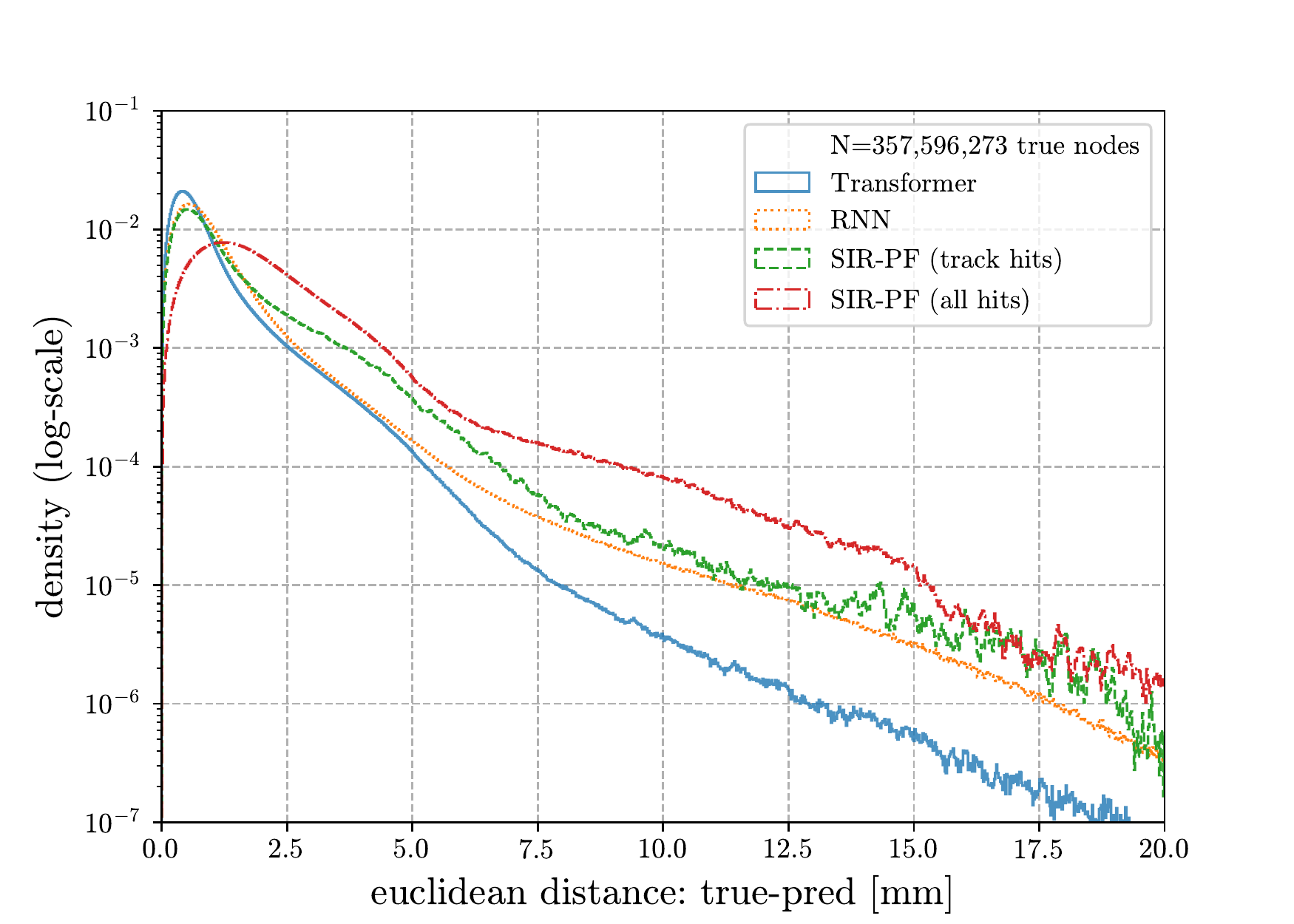}}
    \hfill
    {\adjincludegraphics[height=6.9cm,trim={{.025\width} 0 {.075\width} 0},clip]{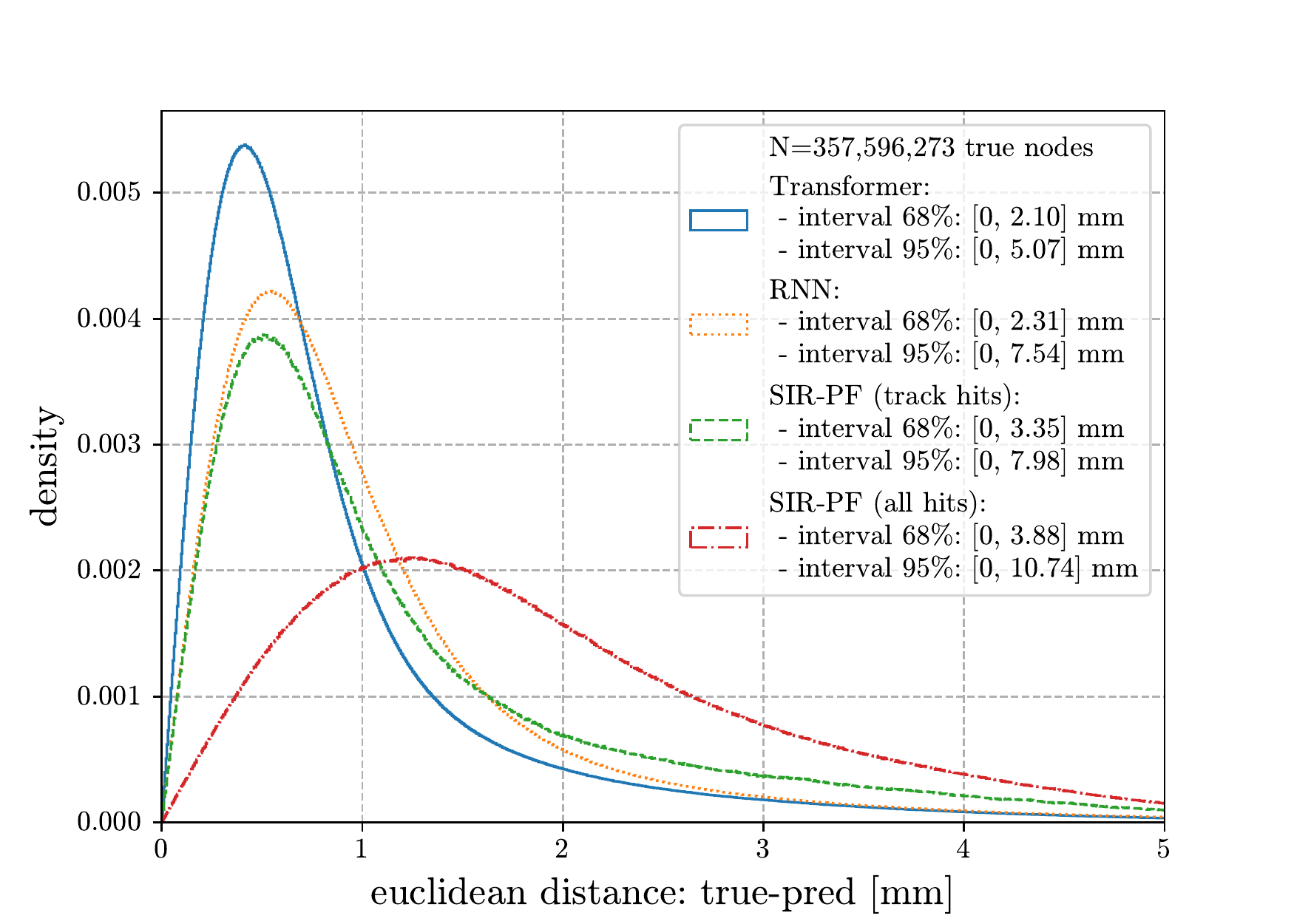}}
\caption{
\label{fig:fitted-nodes}
The distribution of the three-dimensional Euclidean distance between the actual elementary particle position and the corresponding fitted node predicted by the Transformer, the recurrent neural network (RNN), and the Sequential Importance Resampling particle filter (SIR-PF, with only track hits and all hits as input).
The sample used to generate the histograms contains all the simulated particles.
Results show the distributions for a log-scale (left) and normal-scale (right, cropped to a maximum distance of 5 mm) densities, as well as the one-sided area ranges, representing 68\% and 95\% of the distributions.
\label{fig:node-position} 
}
\end{figure*} 

\newcolumntype{g}{>{\columncolor{Gray}}c}
\newcolumntype{h}{>{\columncolor{Gray}}l}
\begin{table*}[htb]
    \centering
    \vspace{0.15cm}
    \resizebox{0.8\linewidth}{!}{%
    \begin{tabular}{c|c|c|gg|gg|gg|gg}
    \hline \hline
    \rule{0pt}{8pt} algorithm & input & particle & \multicolumn{2}{c|}{\CC mean ($\mu$) [mm]} & \multicolumn{2}{c|}{std ($\sigma$) [mm]} & \multicolumn{2}{c|}{68\% area [mm]} & \multicolumn{2}{c}{95\% area [mm]} \\
    \hline
    
    \rule{0pt}{9pt} 
    
    
    & & all & \multicolumn{2}{c}{0.92} & \multicolumn{2}{c}{0.97} & \multicolumn{2}{c}{[0, 2.10]} & \multicolumn{2}{c}{[0, 5.07]} \\
    & & & stopping & escaping & stopping & escaping & stopping & escaping & stopping & escaping \\
    
    \rule{0pt}{9pt} & & $``$ & \CC 1.10  & \CC 0.80 & \CC 1.16 & \CC 0.78  & \CC [0, 2.57] & \CC [0, 1.64] &\CC [0, 5.76] &\CC [0, 4.30] \\
    \multirowcell{2}{\textbf{Transformer}} & \multirowcell{2}{\textbf{all hits}} & $p$ & 1.11 & 0.81 & 1.18 & 0.80 & [0, 2.73] & [0, 1.71] & [0, 5.48] & [0, 4.29] \\
    &  & $\pi^{\pm}$ &\CC 1.07 &\CC 0.83 &\CC 1.09 &\CC 0.78 &\CC [0, 2.40] &\CC [0, 1.70] &\CC [0, 5.63] &\CC [0, 4.22] \\
    & & $\mu^{\pm}$ & 0.94 & 0.71 & 1.04 & 0.66 & [0, 1.80] & [0, 1.33] & [0, 4.57] & [0, 3.81] \\
    && $e^{\pm}$  &\CC 1.18 &\CC 1.07 &\CC 1.23 &\CC 1.06 &\CC [0, 2.74] &\CC [0, 2.44] &\CC [0, 6.81] &\CC [0, 5.36] \\
    
    \hline
    
    
    & & all & \multicolumn{2}{c}{1.13} & \multicolumn{2}{c}{1.25} & \multicolumn{2}{c}{[0, 2.31]} & \multicolumn{2}{c}{[0, 7.54]} \\
    & & & stopping & escaping & stopping & escaping & stopping & escaping & stopping & escaping \\
    
    \rule{0pt}{9pt} & & $``$ & \CC 1.27  & \CC 1.03 & \CC 1.50 & \CC 1.02  & \CC [0, 2.79] & \CC [0, 1.96] & \CC [0, \phantom{1}9.03] & \CC [0, 5.95] \\
    \multirowcell{2}{\textbf{RNN}} & \multirowcell{2}{\textbf{all hits}} & $p$ & 1.26 & 0.99 & 1.48 & 0.98 & [0, 2.92] & [0, 1.86] & [0, \phantom{1}9,52] & [0, 5.48] \\
    &  & $\pi^{\pm}$ &\CC 1.20 &\CC 1.04 &\CC 1.35 &\CC 0.99 &\CC [0, 2.47] &\CC [0, 1.94] &\CC [0, \phantom{1}8.14] &\CC [0, 5.57] \\
    & & $\mu^{\pm}$ & 1.12 & 0.95 & 2.48 & 0.90 & [0, 2.20] & [0, 1.72] & [0, 12.49] & [0, 5.15] \\
    && $e^{\pm}$  &\CC 1.45 &\CC 1.35 &\CC 1.51 &\CC 1.36 &\CC [0, 3.16] &\CC [0, 2.84] &\CC [0, \phantom{1}9.06] &\CC [0, 8.02] \\
    
    \hline
    
    
    & & all & \multicolumn{2}{c}{1.40} & \multicolumn{2}{c}{1.50} & \multicolumn{2}{c}{[0, 3.35]} & \multicolumn{2}{c}{[0, 7.98]} \\
    & & & stopping & escaping & stopping & escaping & stopping & escaping & stopping & escaping \\
    
    \rule{0pt}{9pt} & & $``$ & \CC 1.53  & \CC 1.30 & \CC 1.69 & \CC 1.35  & \CC [0, 3.70] & \CC [0, 3.08] &\CC [0, \phantom{1}9.28] &\CC [0, 6.91] \\
    & \multirowcell{2}{\textbf{track hits}} & $p$ & 1.38 & 1.19 & 1.39 & 1.16 & [0, 3.45] & [0, 2.98] & [0, \phantom{1}6.36] & [0, 5.49] \\
    &  & $\pi^{\pm}$ &\CC 1.46 &\CC 1.29 &\CC 1.74 &\CC 1.26 &\CC [0, 3.59] &\CC [0, 3.04] &\CC [0, \phantom{1}9.45] &\CC [0, 6.05] \\
    & & $\mu^{\pm}$  & 1.24 & 1.19 & 1.22 & 1.22 & [0, 2.76] & [0, 2.73] & [0, \phantom{1}5.87] & [0, 6.25] \\
    \multirowcell{2}{\textbf{SIR-PF}} && $e^{\pm}$  &\CC 1.92 &\CC 1.79 &\CC 1.99 &\CC 1.78 &\CC [0, 4.29] &\CC [0, 3.94] &\CC [0, 11.85] &\CC [0, 9.87] \\
    
    \cline{3-11}
     
    
    & & all & \multicolumn{2}{c}{2.21} & \multicolumn{2}{c}{2.00} & \multicolumn{2}{c}{[0, 3.88]} & \multicolumn{2}{c}{[0, 10.74]} \\
    & & & stopping & escaping & stopping & escaping & stopping & escaping & stopping & escaping \\
    
    \rule{0pt}{9pt} & & $``$ & \CC 2.33  & \CC 2.13 & \CC 2.34 & \CC 1.72  & \CC [0, 4.19] & \CC [0, 3.68] &\CC [0, 12.30] &\CC [0, \phantom{1}9.54] \\
    & \multirowcell{2}{\textbf{all hits}}& $p$  & 2.33 & 2.14 & 2.21 & 1.83 & [0, 4.33] & [0, 3.84] & [0, 12.37] & [0, 10.08] \\
    &  & $\pi^{\pm}$ &\CC 2.23 &\CC 2.15 &\CC 2.35 &\CC 1.72 &\CC [0, 3.90] &\CC [0, 3.73] &\CC [0, 11.80] &\CC [0, \phantom{1}9.30] \\
    & & $\mu^{\pm}$ & 2.18 & 2.03 & 3.53 & 1.56 & [0, 3.82] & [0, 3.41] & [0, 21.26] & [0, \phantom{1}8.57] \\
    && $e^{\pm}$  &\CC 2.51 &\CC 2.50 &\CC 2.24 &\CC 2.16 &\CC [0, 4.59] &\CC [0, 4.63] &\CC [0, 12.43] &\CC [0, 11.37] \\
    
    \hline \hline
    \end{tabular}}

    \caption{Euclidean distance (mean $\mu$, standard deviation $\sigma$, and ranges for the one-sided 68\% and 95\% areas) between the predicted and the true nodes for the Transformer, recurrent neural network (RNN), and the Sequential Importance Resampling particle filter (SIR-PF) algorithms. For the latter, the table shows the results after inputting: (1) all the hits, (2) track hits only. It also shows the results independently for each particle type (proton $p$, pion $\pi^{\pm}$, muon $\mu^{\pm}$, and electron $e^{\pm}$) and distinguishes whether the particle escaped or stopped at the detector.}
    \label{tab:results_resolution}
\end{table*}

For the SIR-PF, we have considered two different scenarios that vary in the reconstructed input information to the filter: (1) all the reconstructed 3D hits are used as input; (2) only real track hits\footnote{With ``real track hits'' we refer to hits from cubes the actual particle has passed through.} are used as input, which is unavailable information for actual data (and represents a nonphysical scenario) but allows us to test the ideal performance for the current filter. The input for the RNN and Transformer always consisted of all the reconstructed 3D hits. Figure~\ref{fig:node-position} shows a comparison of the performance for the three methods (considering the SIR-PF variant with all the reconstructed hits as input). The results indicate that the Transformer outperforms the other techniques (even for the case with only track hits). Besides, the RNN reports significantly better results than the SIR-PF with only track hits used as input and slightly better fittings concerning the SIR-PF with all hits used as input, which demonstrates not only that the NN-based approaches can handle crosstalk hits but also go beyond and accomplish spatial determination $<$1.5 mm far (on average) from the real physical case.

\begin{figure*}[htb]
  { 
    \adjincludegraphics[height=5.55cm,trim={2.75cm 2.5cm 3.5cm
    4cm},clip]{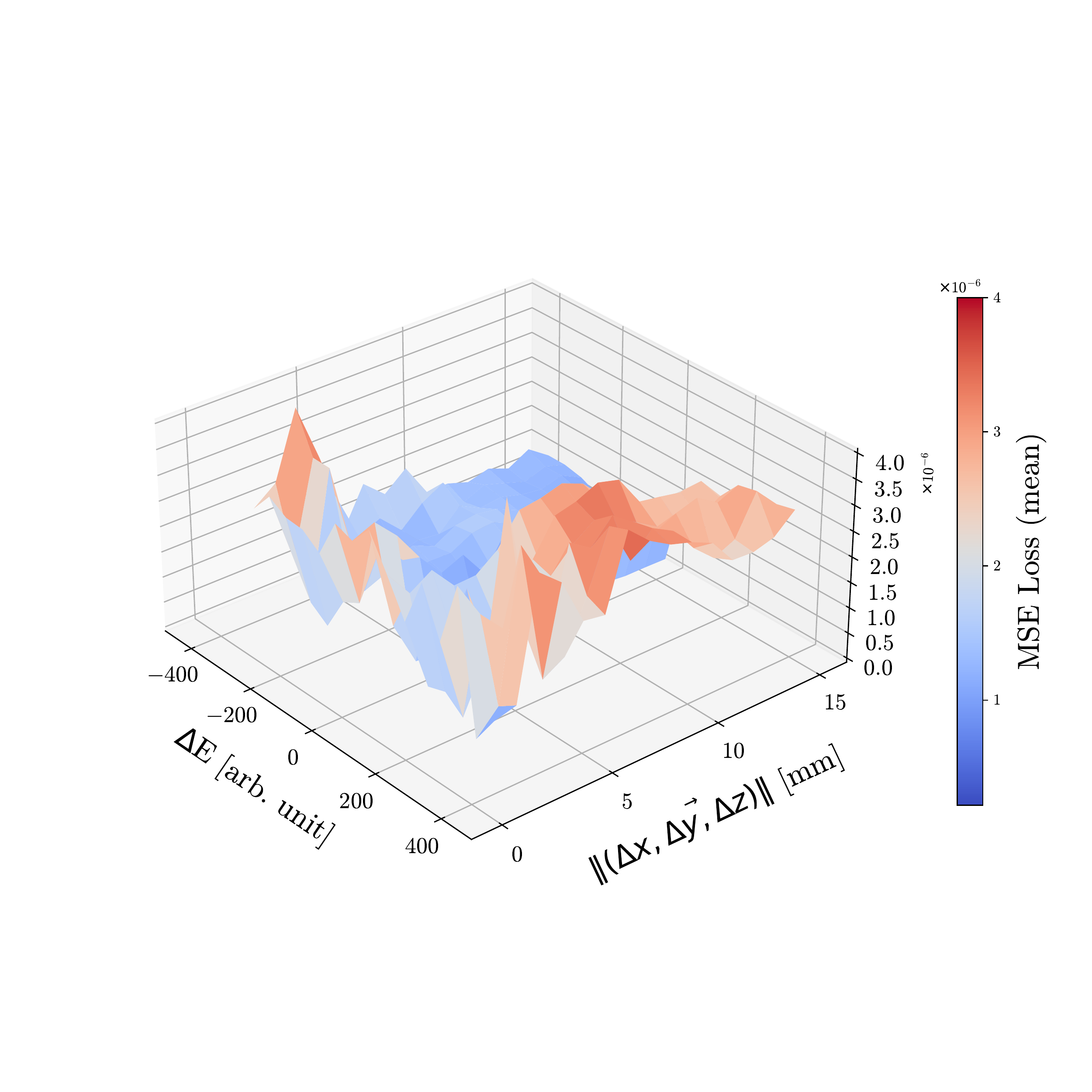}
    \adjincludegraphics[height=5.55cm,trim={2.75cm 2.5cm 3.5cm 4cm},clip]{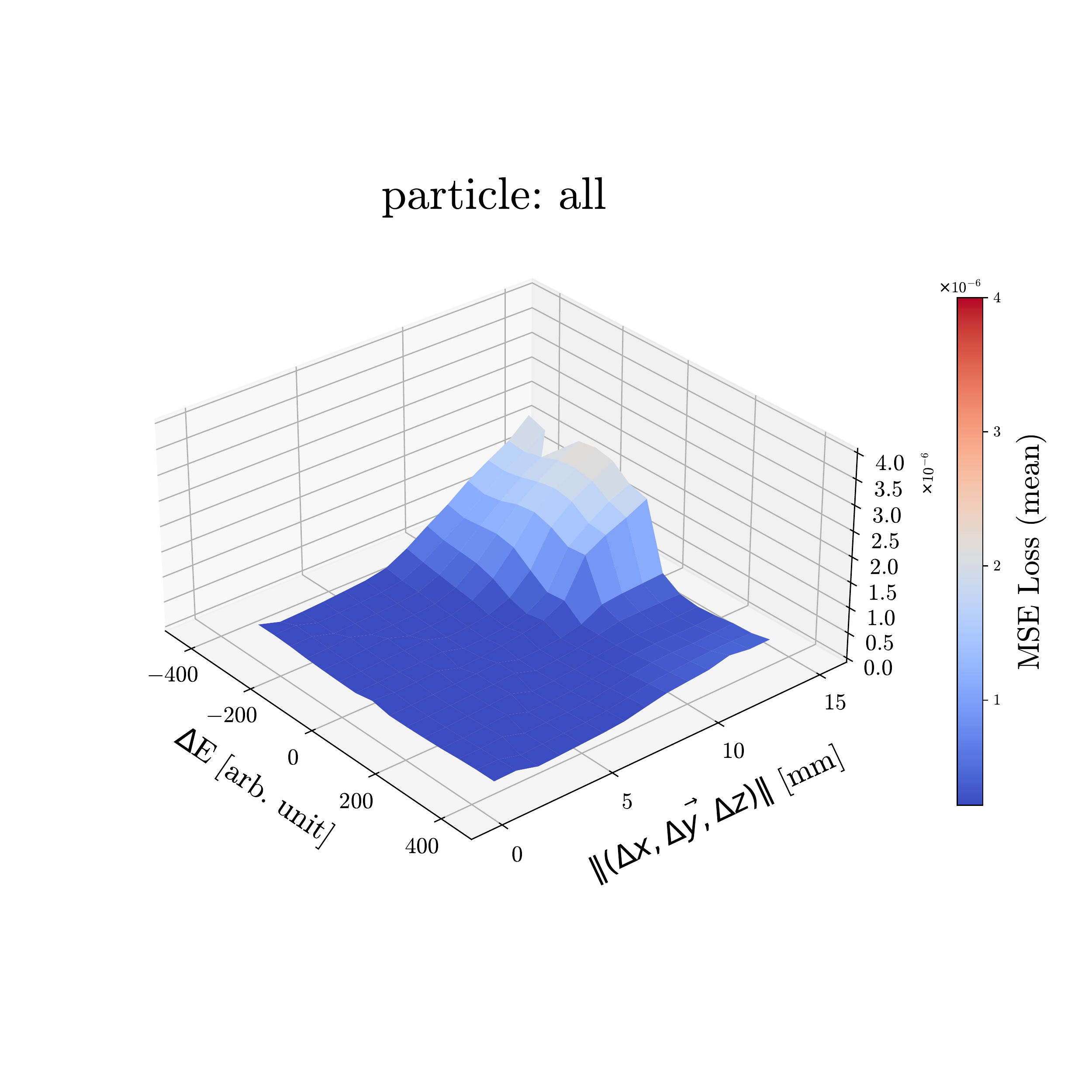}
    \adjincludegraphics[height=5.55cm,trim={2.75cm 2.5cm 1.05cm 4cm},clip]{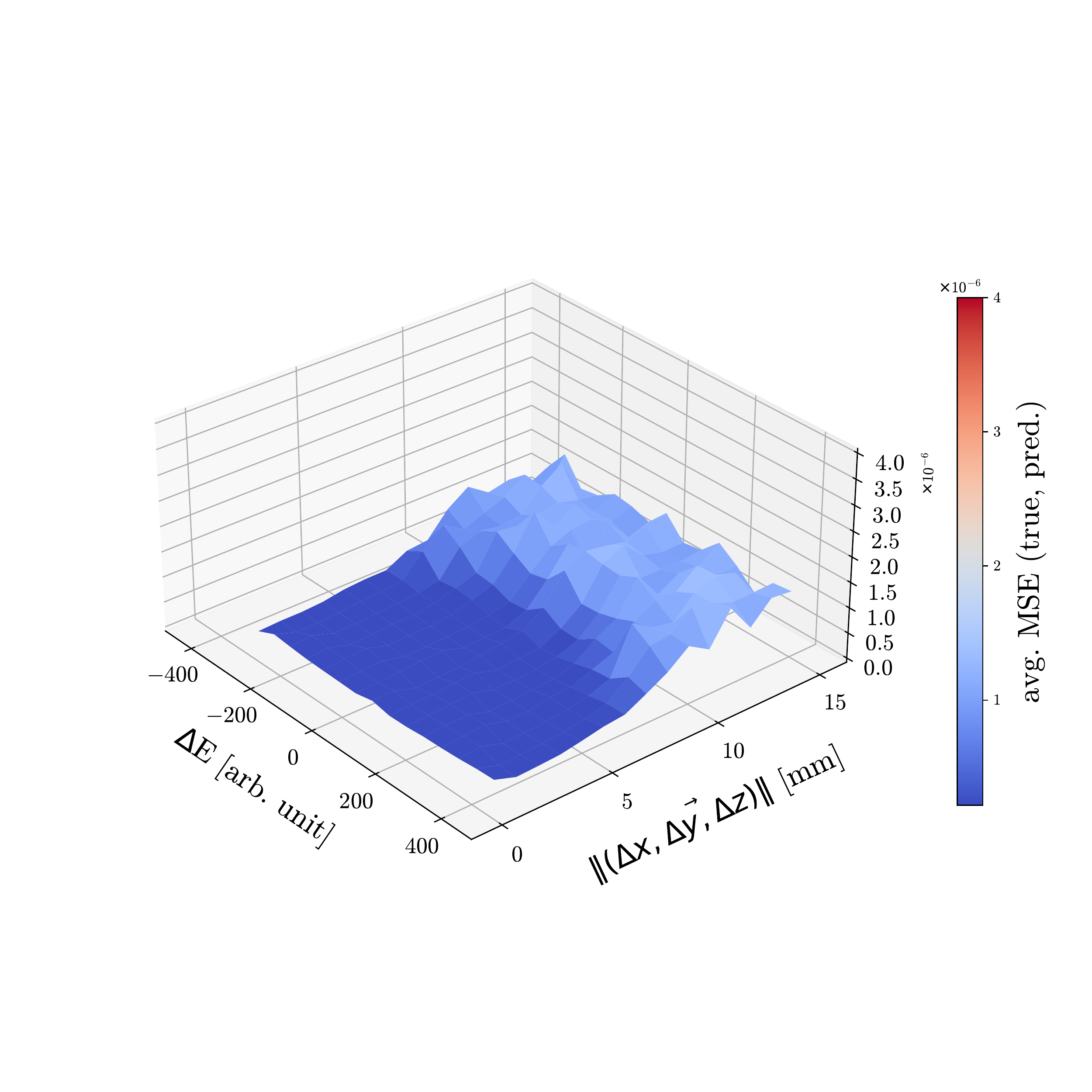}
  }
  { 
    \adjincludegraphics[height=5.55cm,trim={2.75cm 2.5cm 3.5cm 4cm},clip]{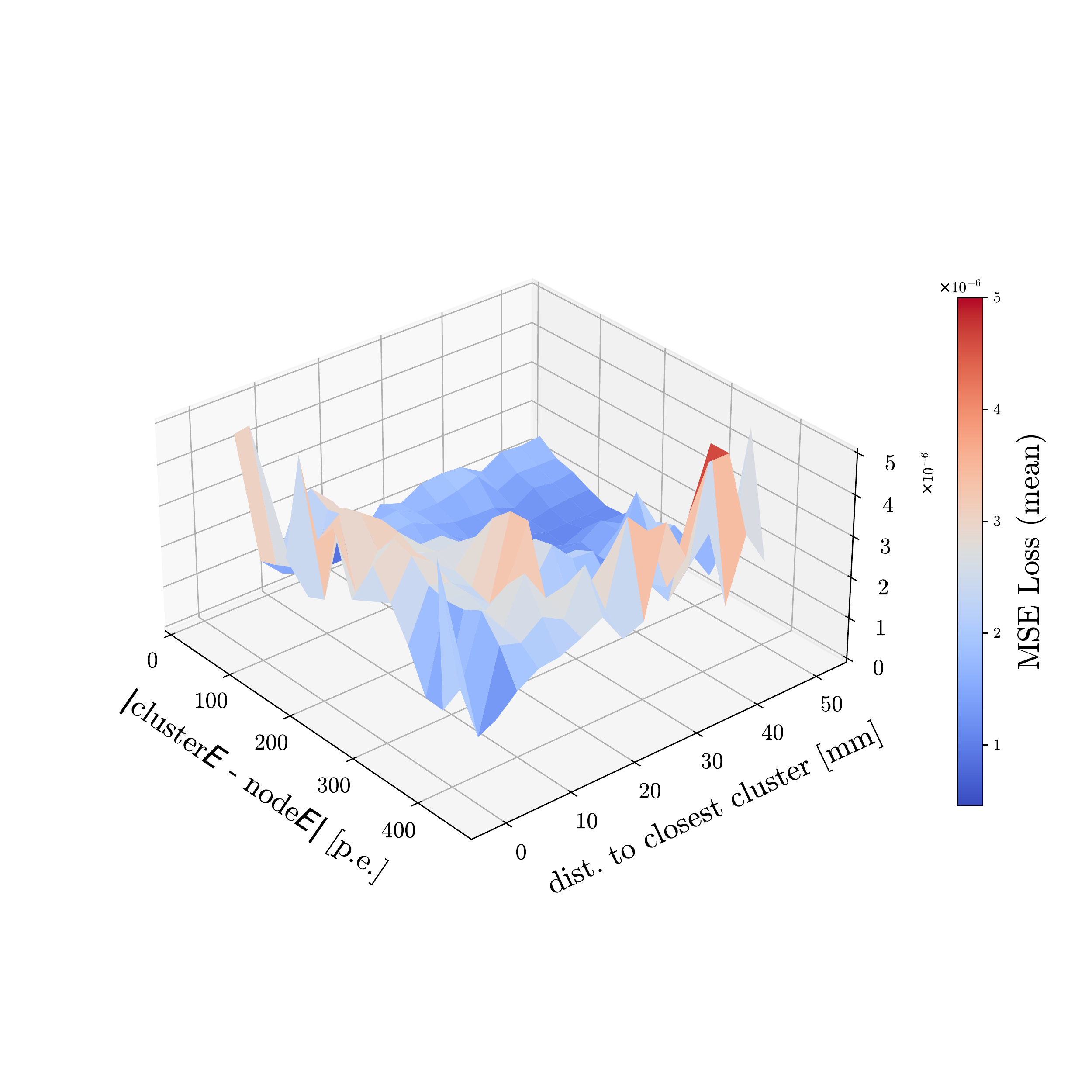}
    \adjincludegraphics[height=5.55cm,trim={2.75cm 2.5cm 3.5cm 4cm},clip]{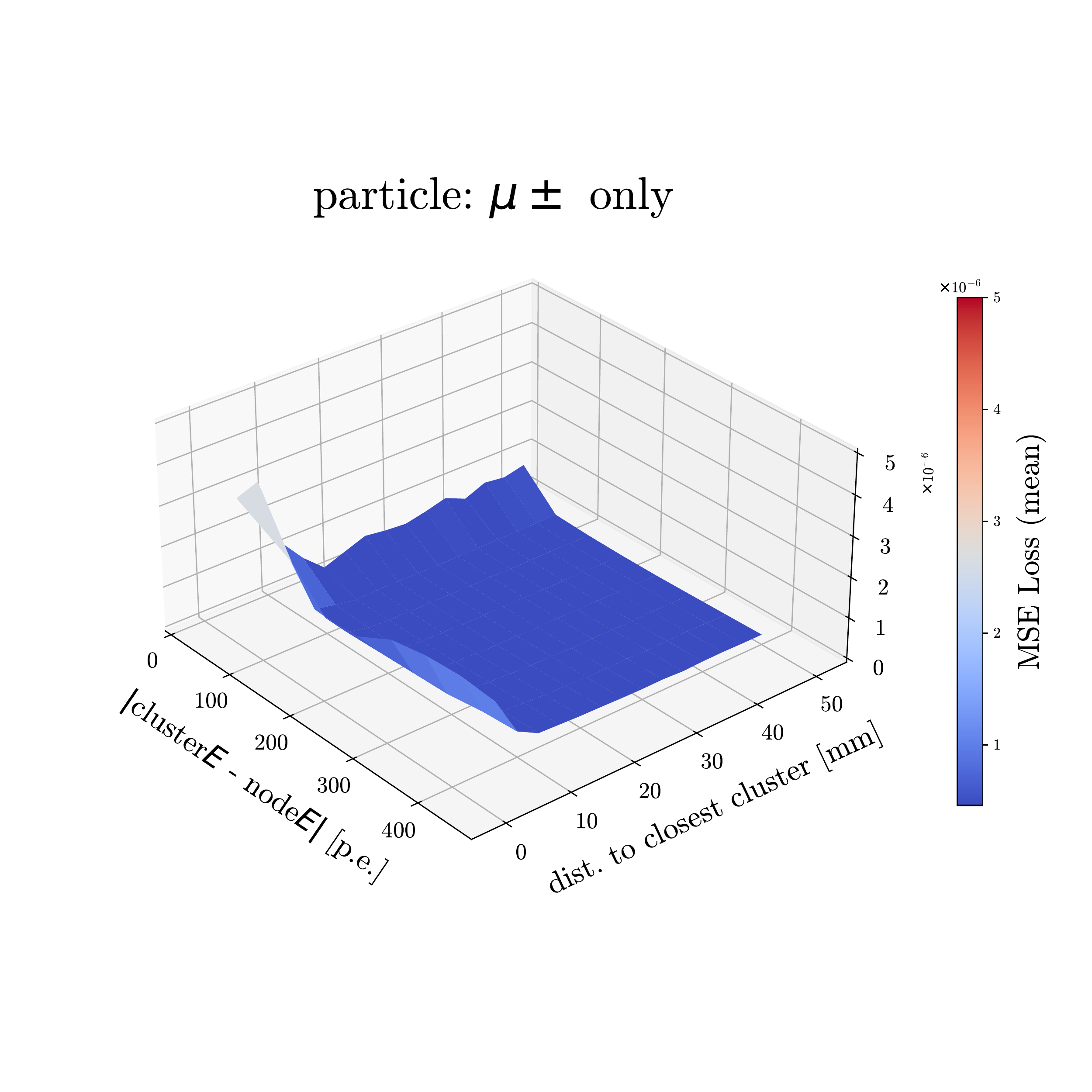}
    \adjincludegraphics[height=5.55cm,trim={2.75cm 2.5cm 1.05cm 4cm},clip]{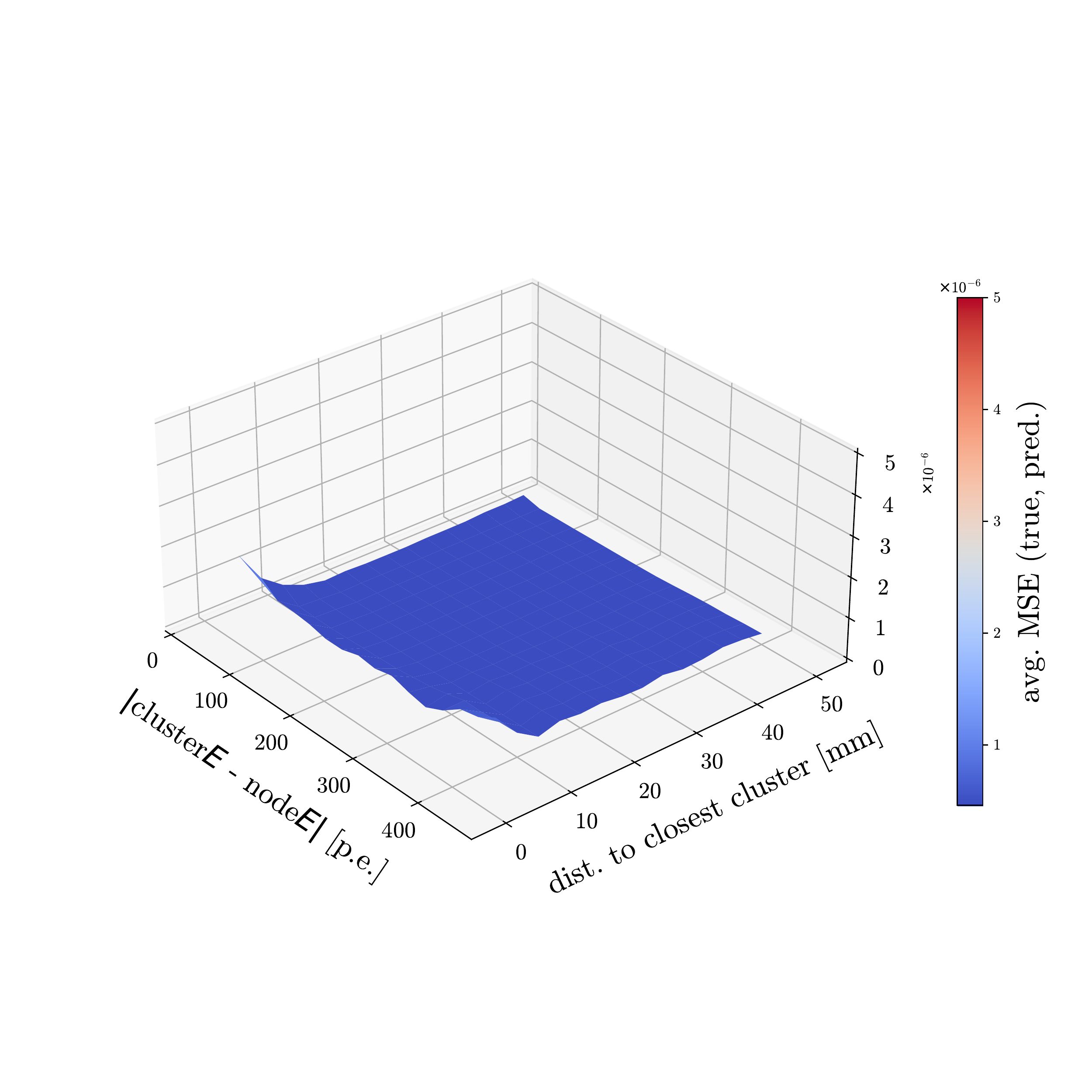}
  }
\caption{
 (Top) Behaviour of the mean-squared-error (MSE) loss concerning $\lVert \vec{(\Delta x, \Delta y, \Delta z)} \rVert$ (magnitude of the vector resulting from the differences in position between consecutive nodes) and $\Delta E$ (differences in energy deposition between successive nodes) for the three algorithms: Sequential Importance Resampling particle filter (SIR-PF) with all hits (left), recurrent neural network (RNN, middle), and Transformer (right). After standardisation, each bin corresponds to the average mean-squared error (MSE) loss applied to the pair (true node, fitted/predicted node). All fitted nodes are considered. (Bottom) Behaviour of the mean-squared-error (MSE) loss concerning the distance from each fitted node to the closest cluster hit and $|clusterE-nodeE|$ (absolute difference between the energy depositions of the fitted node and the nearest cluster hit) for the three algorithms: SIR-PF with all hits (left), RNN (middle), and Transformer (right). After standardisation, each bin corresponds to the average mean-squared error (MSE) loss applied to the pair (true node, fitted/predicted node). Only nodes from muon ($\mu^{\pm}$) particles are considered.
\label{fig:3dlosses} }
\end{figure*} 

A more exhaustive analysis of the performance of both methods is presented in Tab.~\ref{tab:results_resolution}, which reveals the effectiveness of the NNs compared to the SIR-PF variants. The table also confirms that the track fitting becomes more manageable when the crosstalk hits are removed from the input and more precise information is given to the filter (the SIR-PF version with only track hits outperforms the one with all hits as input). This last fact also evidences the power of deep learning, which is, on average, able to predict more accurately the node positions and thus the true track trajectory, even if its input consists of all the reconstructed hits without any type of pre-processing (e.g., removal of crosstalk hits), meaning that it could understand the relations between hits internally, confirming the ability to discard the crosstalk hits during the fitting calculation. In order to compare the Transformer and the RNN, it is worth looking at the muon fitting at Tab.~\ref{tab:results_resolution}: the Transformer reports the best results for fitting muon particles (for both mean and standard deviation) in contrast to the RNN, which reports an atypically large std dev. for muon tracks contained in the detector. The explanation relies on the length of the particles and the properties of the algorithms: since muons tend to have the most extended track length among the simulated stopping particles (protons and pions tend to have more secondary interactions and electrons produce electromagnetic showers), and the RNN depends on its memory mechanisms to bring features from faraway hits to fit a particular reconstructed hit (see Sec.\ref{sec:supplementary_info}, Supplementary Information, for more details), it is habitual to omit some information from remote hits during the fitting; on the other hand, the Transformer reduces its mistakes by having a complete picture of the particle thanks to its capacity to learn the correlations among all reconstructed hits.

To understand the behaviour of the fittings for the different physical structures of the particles, we have calculated the mean-squared error (MSE, which is the loss function used during the neural network trainings) between each fitted and true node and visualised the information in Fig.~\ref{fig:3dlosses}. The MSE loss, which penalises outliers by construction, seems flatter for the RNN and Transformer than for the SIR-PF, indicating more stability in the fitting. Besides, it is notorious for highlighting the tendency for particular negative $\Delta E$ values to report high losses in the NN cases, caused mainly due to the low charge of crosstalk compared to track hits. Besides, Fig.~\ref{fig:3dlosses}, as expected, also reveals that the three algorithms report worse fittings when getting closer to cluster hits connected to the track. For instance, in the case of muon particles, these clusters are typically due to the ejection of $\delta$-rays, 
i.e. orbiting electrons knocked out of atoms, often causing a kink on the muon track; 
however, both NNs seem to deal much better with this attribute.

Even if the primary goal of this article is to show the performance of the fitting from a physics perspective, it is worth comparing the different algorithms in terms of computing time. Table~\ref{tab:computing_time} manifests the average time it takes for each algorithm to run the fitting on a single particle. The results exhibit a considerable speedup for both the RNN and the Transformer models (with speedups of $\sim\times $4 and $\sim\times$35, respectively) with a single thread on the CPU. The table does not show the SIR-PF results for the distributed computing scenarios since it would require some time to parallelise the SIR-PF code to run it with multiple threads or to adapt it to GPU computation, which is clearly beyond the scope of the study; that being said, the table shows the parallel results for the RNN and Transformer cases since these are features available in the PyTorch framework, which show how inexpensive it would be to achieve significant speedups for an ordinary user.

\begin{table}[htb]
    \centering
    \resizebox{1.0\linewidth}{!}{%
    \begin{tabular}{c|l|crr}
    \hline \hline
    \rule{0pt}{8pt} Processor & Parallelisation & \multicolumn{1}{c}{SIR-PF} & \multicolumn{1}{c}{RNN} & \multicolumn{1}{c}{Transformer} \\
    \hline
    \multirowcell{2}{\textbf{CPU}} & single-thread & $435.71\pm 5.18$ & $91.16\pm 1.17$ & $12.25\pm 0.19$  \\
                                   & multi-thread & - & $82.22\pm 1.00$ & $6.58\pm 0.04$  \\
    \hline
    \multirowcell{3}{\textbf{GPU}} & batch\_size = 1 & - & $31.27\pm 0.99$ & $8.96\pm 0.31$  \\
                                   & batch\_size = 16 & - & $4.02\pm 0.12$ & $1.24\pm 0.12$  \\
                                   & batch\_size = 64 & - & $1.43\pm 0.05$ & $0.71\pm 0.04$  \\
    \hline \hline
    \end{tabular}}

    \caption{Average computing time each algorithm takes to process a single particle (in milliseconds). The test shows the average results of running the three methods (Sequential Importance Resampling particle filter (SIR-PF) with all hits, recurrent neural network (RNN), and Transformer) on the same ten random subsets of the testing dataset consisting of 10,000 particles each. CPU: AMD EPYC 7742 64-Core 3200MHz Processor, GPU: NVIDIA A100 Tensor Core (8GB of memory). Note that the SIR-PF implemented does not support multi-threading nor GPU computation since it is out of the scope of the article; parallelising the computation for the RNN and Transformer becomes trivial thanks to PyTorch. The parameter ``batch\_size'' indicates the number of particles processed together in each step.}
    \label{tab:computing_time}
\end{table}

Finally, if we look at the size of the histogram used to calculate the likelihood, it consists of 3,948,724 bins with non-zero values, compared to the 213,553 learnt parameters of the RNN ($\sim$18 times fewer parameters) and the 167,875 parameters of the Transformer ($\sim$23 times fewer parameters than the SIR-PF histogram). Of course, it would be possible to design a more efficient version of the histogram (which is also out of the scope of the article) to reduce the difference in parameters among the methods. Nevertheless, this first approximation already gives insights into how compact the information is encoded in the neural network cases in contrast to the Bayesian filter scenario with a physics-based likelihood calculation.

\begin{figure*}[htb]
    {\adjincludegraphics[width=8.7cm,trim={{.015\width} 0 {0.00\width} 0},clip]{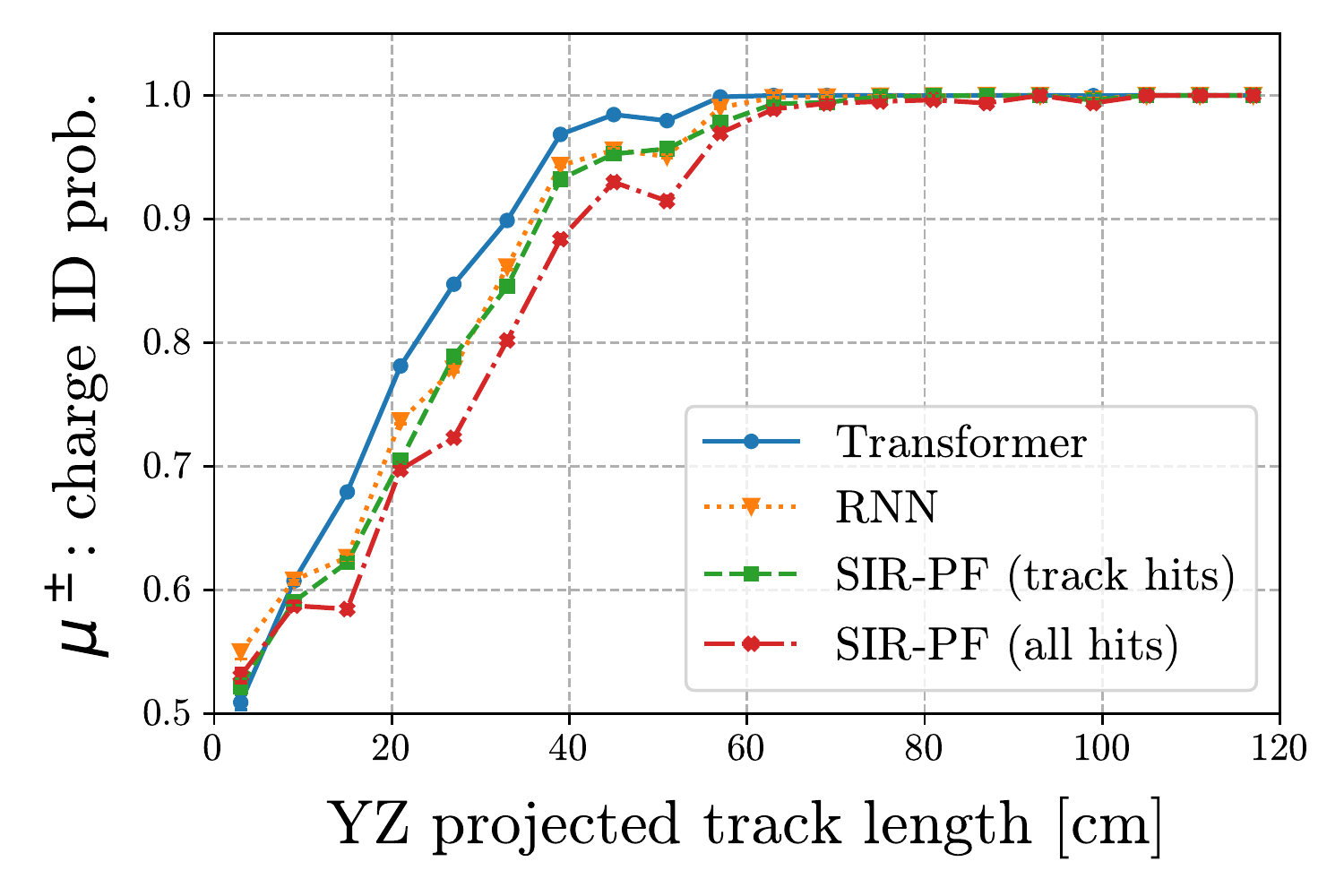}} 
    \hfill
    {\adjincludegraphics[width=8.7cm,trim={{.015\width} 0 {0.00\width} 0},clip]{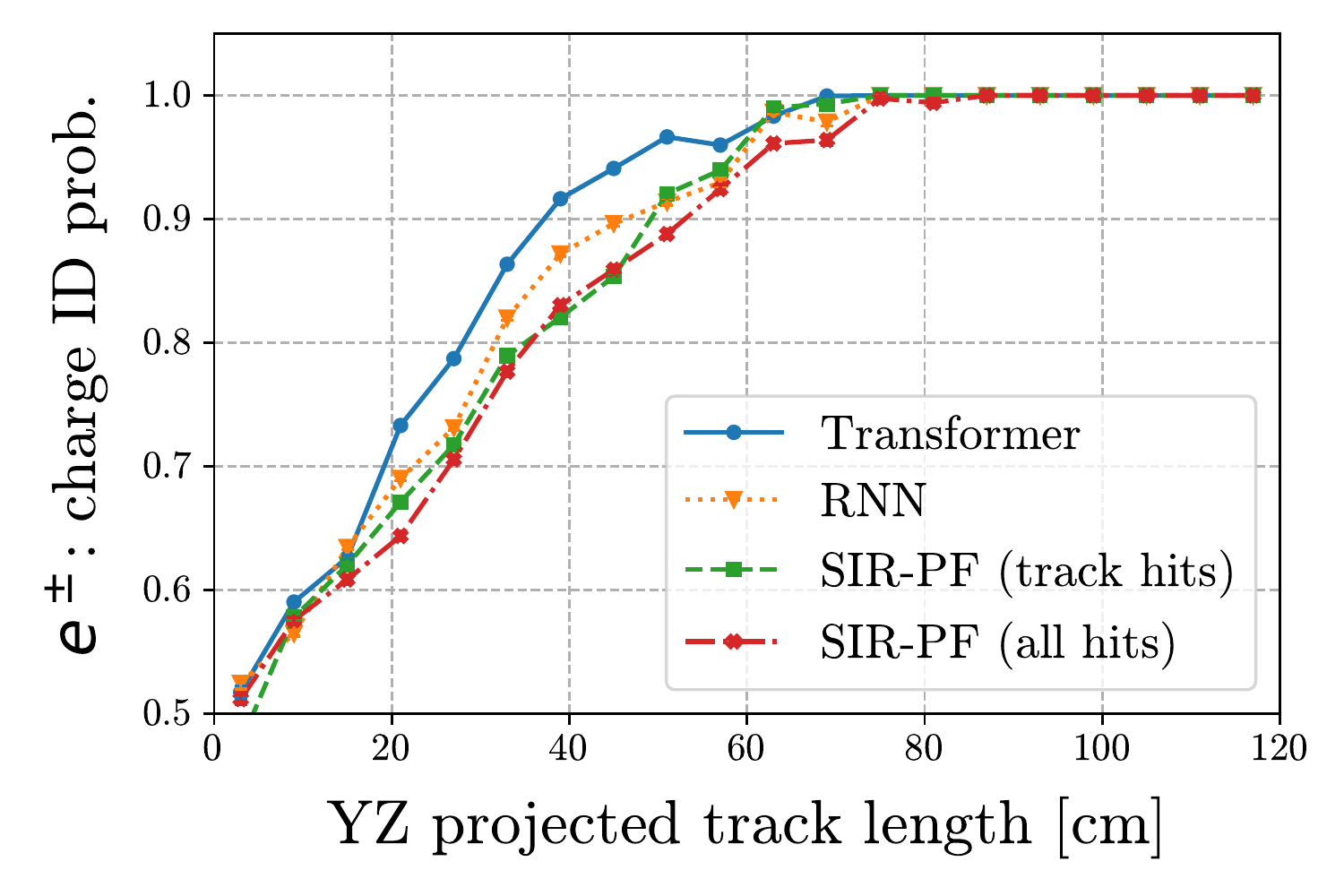}}
    {\adjincludegraphics[width=8.7cm,trim={{.015\width} 0 {.0\width} 0},clip]{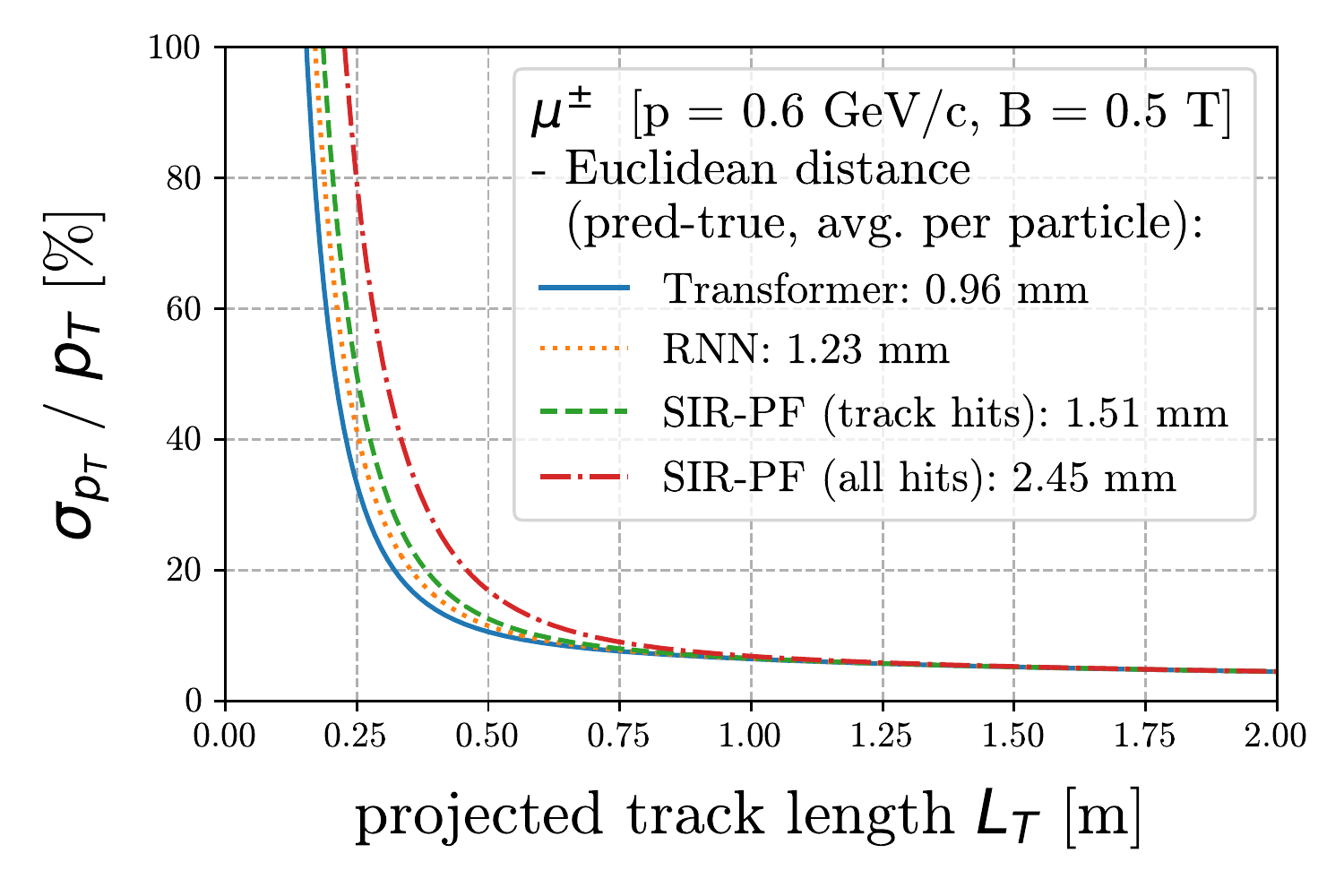}} 
    \hfill
    {\adjincludegraphics[width=8.7cm,trim={{.015\width} 0 {.0\width} 0},clip]{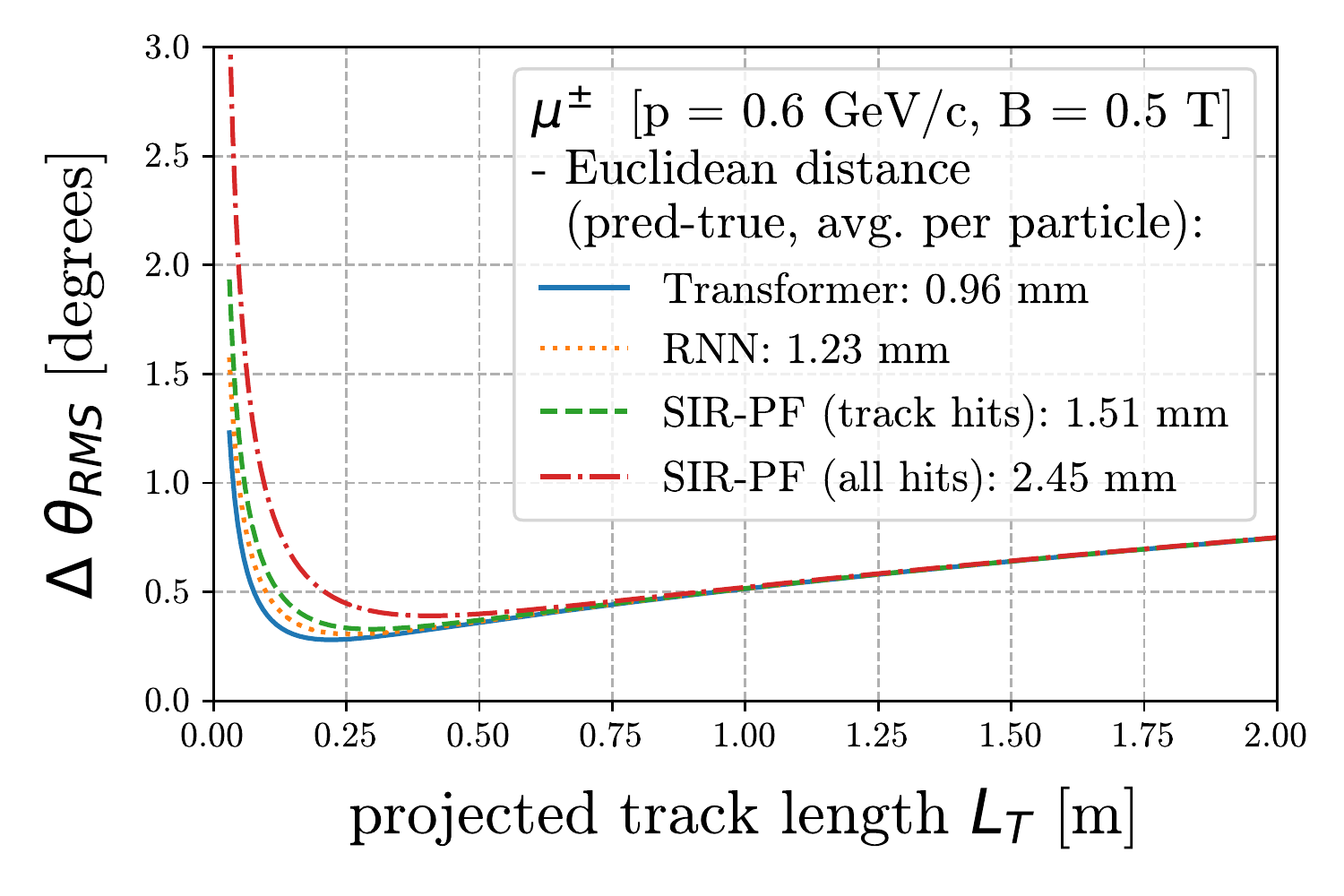}}
\caption{
\label{fig:chargeID_len_muon} 
(Top) Charge ID probability for muons and antimuons ($\mu^{\pm}$, left) and electrons and positrons ($e^{\pm}$, right) as a function of the track length projected on the plane perpendicular to the 0.5~T magnetic field. An equal number of particles and antiparticles are considered in both cases.
\label{fig:momentum-angle-resolution} 
(Bottom) A muon example of 0.6~GeV/c with a 0.5~T magnetic field is considered to show the momentum-by-curvature resolution as a function of the track length projected on the plane perpendicular to the magnetic field (left), as well as the angular resolution as a function of the particle length in the detector. The average Euclidean distance (between true and fitted nodes) per muon particle was considered, and the results are presented for the different fitting techniques: Transformer, recurrent neural network (RNN), and Sequential Importance Resampling particle filter (SIR-PF) with all hits and only track hits as input.
}
\end{figure*}

\review{\subsection{Impact on the detector physics performance}}
\label{sec:performance-detector}

The reconstruction of the primary particle kinematics provides diverse information: the electric charge (negative or positive); 
the identification of the particle type (protons, pions, muons, electrons), which mainly depends on the particle stopping power as a function of its momentum; the momentum, either from the track range of the particle that stops and releases all its energy in the detector active volume or from the curvature of its track if the detector is immersed in a magnetic volume;
the direction.
%
An improved resolution on the spatial coordinate and, consequently, of the particle stopping power impacts the accuracy and precision of the physics measurement.
This section compares the performance of the reconstruction of particle interactions provided by the Transformer and RNN to the one using the SIR-PF.


\begin{figure*}[htb]
    {\adjincludegraphics[height=6.1cm,trim={{.025\width} 0 {.05\width} 0},clip]{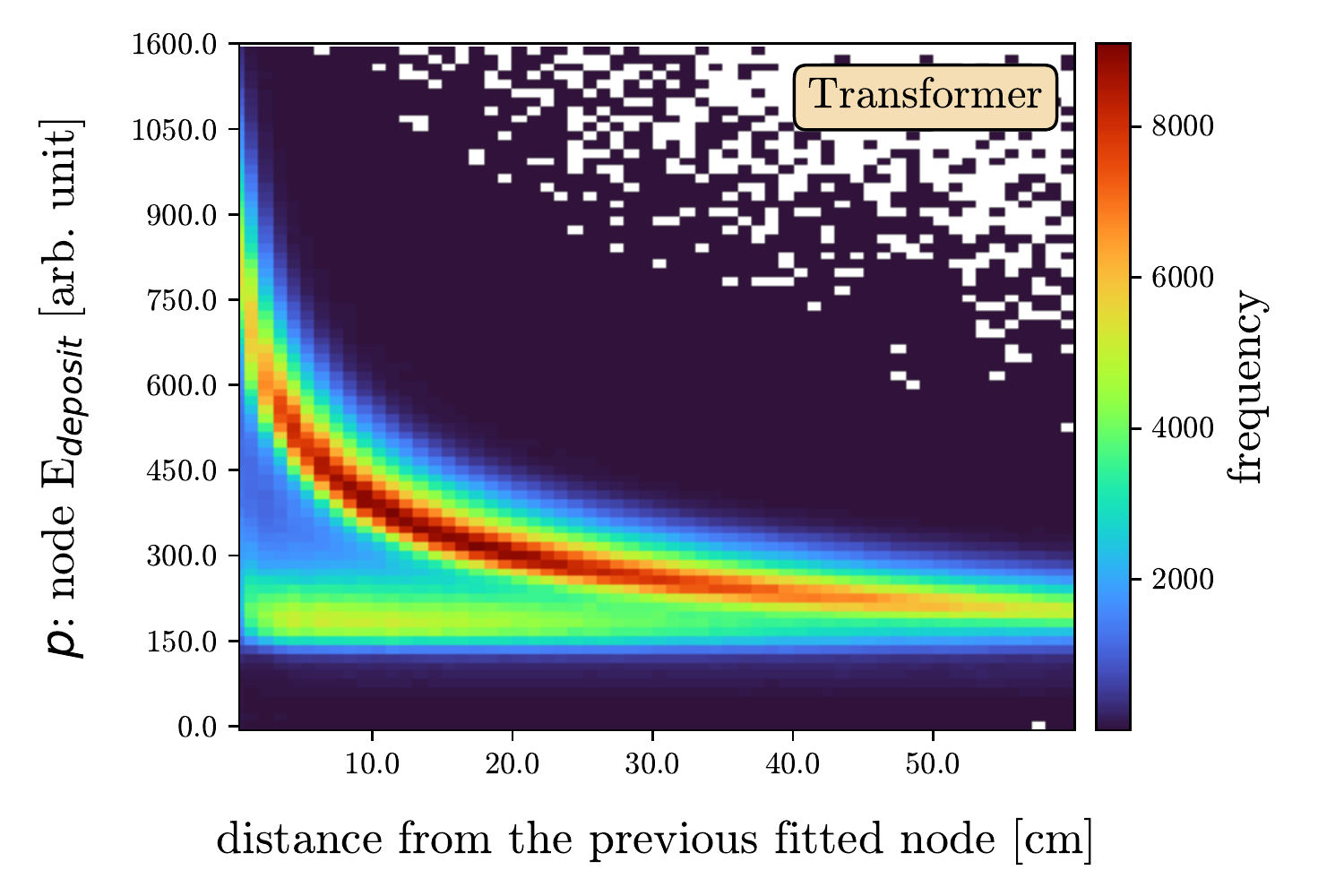}}
    \hfill
    {\adjincludegraphics[height=6.1cm,trim={{.025\width} 0 {.05\width} 0},clip]{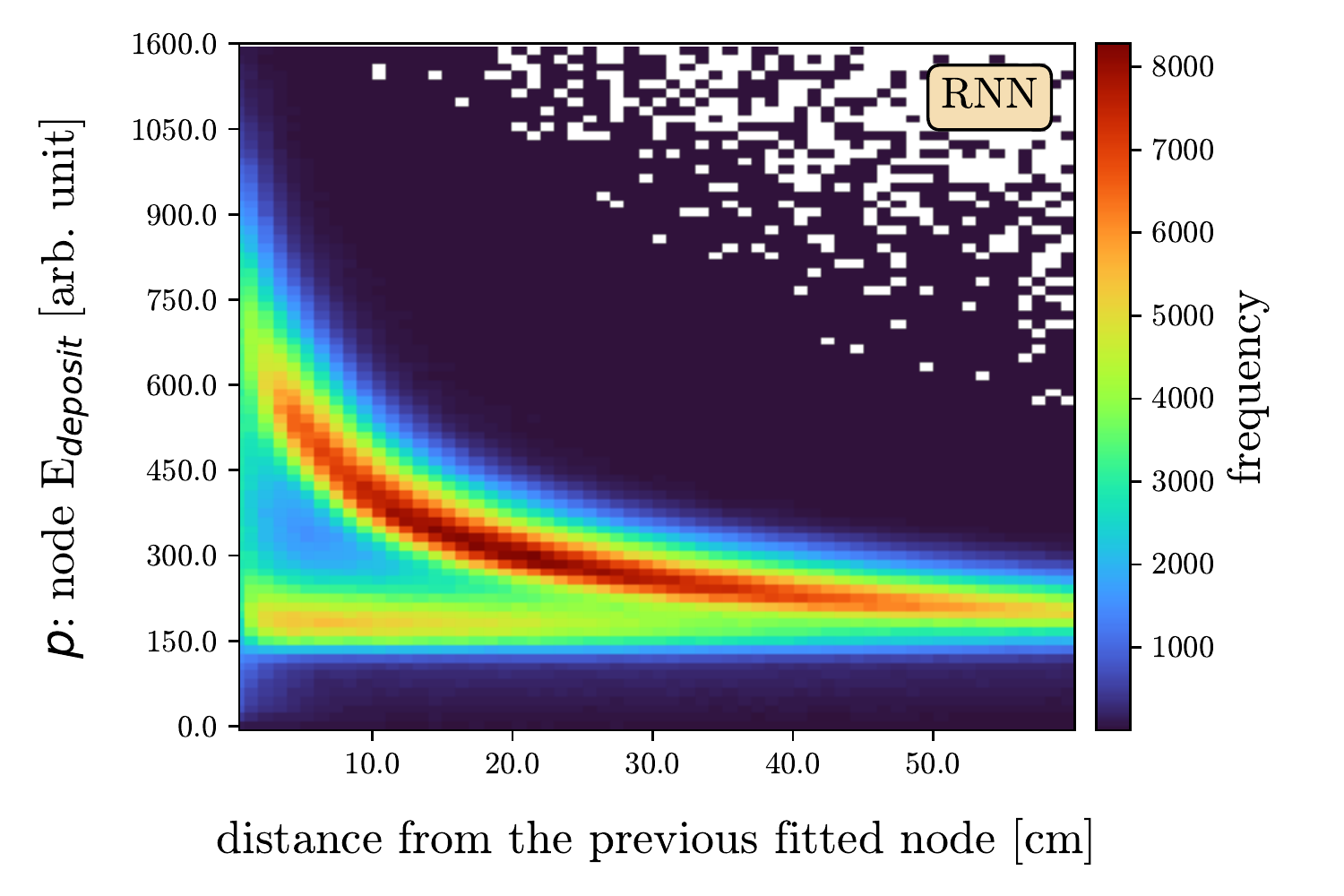}}
    %
    %
    {\adjincludegraphics[height=6.1cm,trim={{.025\width} 0 {.05\width} 0},clip]{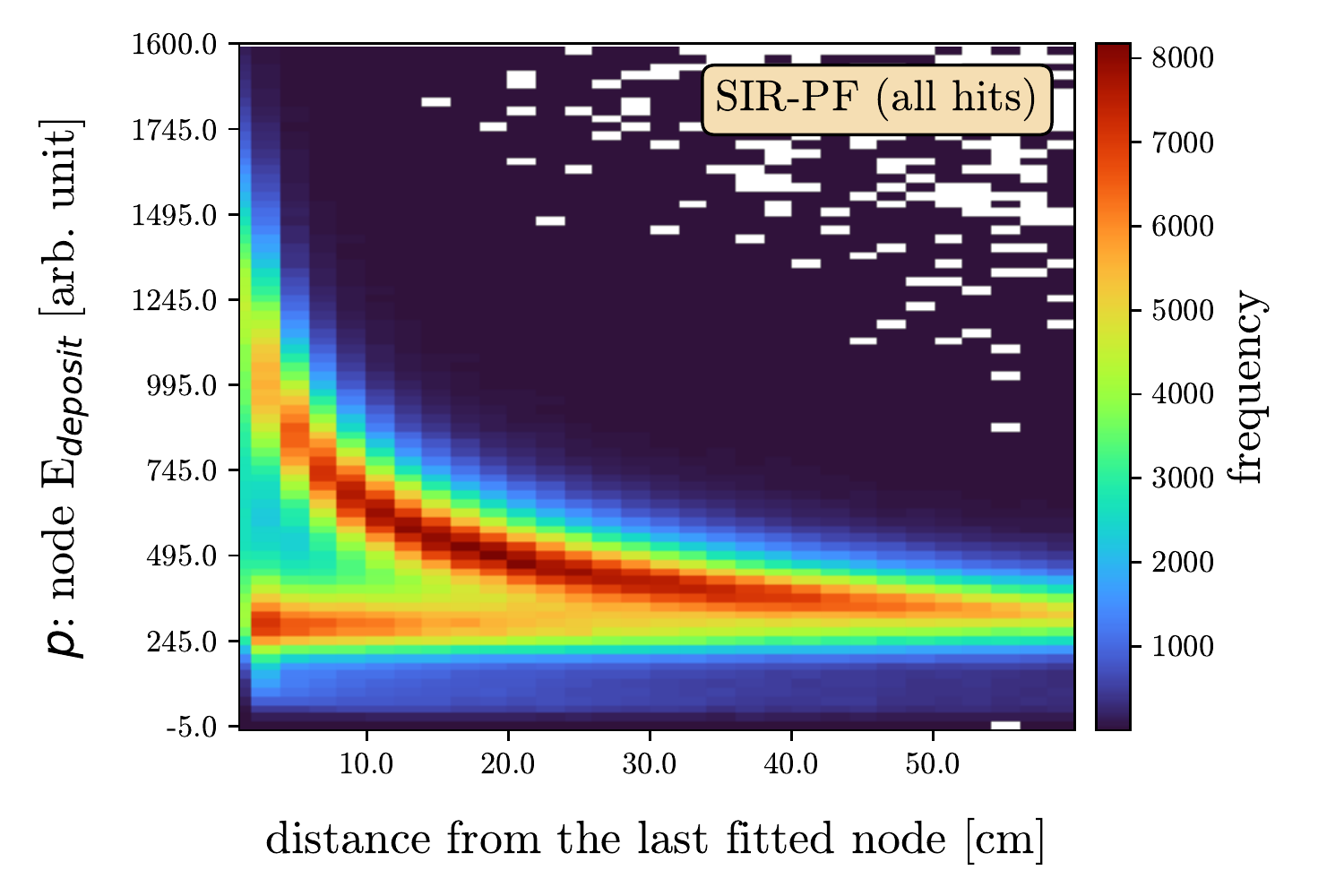}}
    \hfill
    {\adjincludegraphics[height=6.1cm,trim={{.025\width} 0 {-0.035\width} 0},clip]{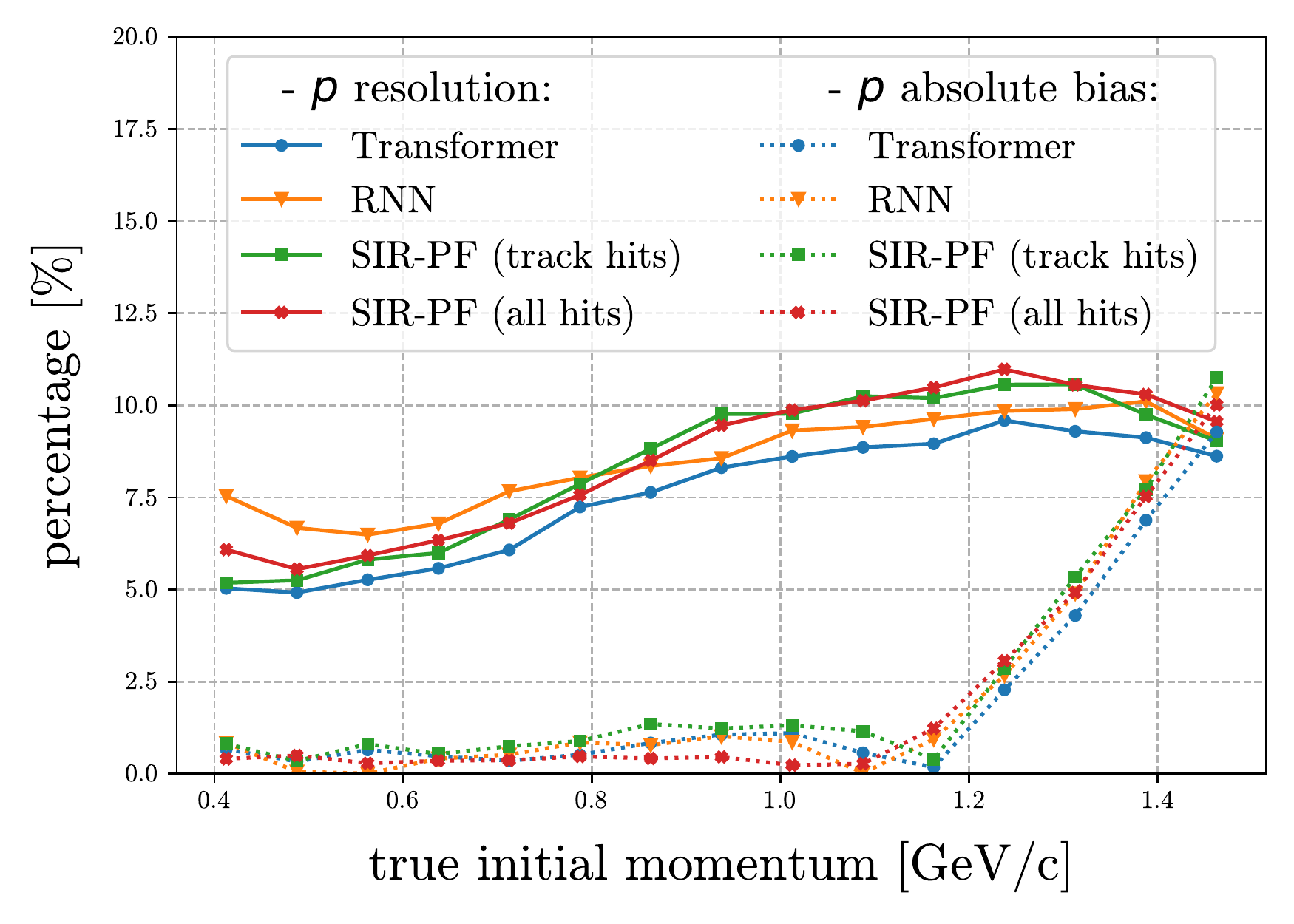}} 
    \caption{
    \label{fig:dedx_proton_RNN} 
    Measured energy deposited by a stopping proton at each fitted node as a function of its distance from the last fitted node for the Transformer (top left), the recurrent neural network (RNN, top right), and Sequential Importance Resampling particle filter (SIR-PF) with all hits as input (bottom left). Note that we chose a different binning for the SIR-PF than the one used for the NN versions for visualisation reasons since the former algorithm reports fewer fitted nodes per particle on average.
    \label{fig:diff_comb_len_bdtg} 
    (Bottom right) The reconstructed momentum
    bias (dashed line) and resolution (solid line) for stopping protons as a function of real initial proton momentum are shown for the different fitting algorithms.     }
\end{figure*}

The charge identification (charge ID) is performed by reconstructing the curvature of the particle track in the detector immersed in the 0.5~T magnetic field. 
%
The charge ID performance was studied for muons (resp. electrons) with momenta between
0 and 2.5 GeV/c (resp. 0 and 3.5 GeV/c) and isotropic direction distribution. 
From Fig.~\ref{fig:chargeID_len_muon}, it is evident that the NNs outperform the SIR-PF. For instance, the muon charge can be identified with an accuracy better than 90\% if the track has a length projected on the plane transverse to the magnetic field of $\sim$33 and $\sim$36~cm for the Transformer and RNN, respectively.
Instead, the SIR-PF (with all the hits, the version with the same input as the neural network cases) requires a track of at least $\sim$42~cm in order to achieve the same performance.
Similar conclusions can be derived from the charge ID study on electrons and positrons.

In Fig.~\ref{fig:chargeID_len_muon}, the case of a 0.6~GeV/c muon 
was also studied, showing the node positions fitted with the NNs and SIR-PF, with the Transformer better capturing the curvature due to the magnetic field. 
It was found that if the tracking resolution is accurate, it is possible to either improve the detector performance beyond its design or to aim for a more compact design of the scintillator detector deployed in a magnetic field.
For instance, the spatial resolution achieved with the NNs in a magnetic field of 0.5~T allows measuring the momentum of a 0.6~GeV/c muon from its curvature with a resolution of about 15\% with a length of the track projected on the plane transverse to the magnetic field of almost 40~cm, shorter by about 20~cm than the length needed by the SIR-PF with all the hitsO.
Such an improvement implies the possibility of accurately reconstructing the momentum of muons escaping the detector for a larger sample of data.
At the same time, improved methods for the reconstruction of particle interactions could become a new tool in the design of future particle physics experiments, for example leading to more compact detectors, thus lower costs.
Similar conclusions can be achieved about the particle angular resolution, improved by about a factor of two and, simultaneously, requiring a track length three times shorter than the one obtained with traditional methods.

The Transformer outperforms the SIR-PF also in the reconstruction of the particle momentum, both by range and curvature.
For instance, the momentum-by-range resolution for protons stopping in the detector between 0.9 and 1.3~GeV/c is improved by a factor of $\sim$15\%, as shown in Fig.~\ref{fig:diff_comb_len_bdtg}. Since protons typically have a much stronger stopping power towards the end of the track (Bragg peak), the total amount of energy leaked to the adjacent cubes is more significant. We observe that the fitting near the Bragg peak becomes more challenging for protons (for example, compared to muons) and less precise due to the presence of more crosstalk hits. This becomes particularly relevant for low momentum (true initial momentum from 0.4 to 0.8 GeV/c) - hence short - protons. However, the Transformer seems to deal well with this difficulty, whilst the RNN reports worse resolutions for this particular case, as shown in Fig.~\ref{fig:diff_comb_len_bdtg}.

The particle identification performance depends on the capability of reconstructing the particle 
stopping power 
along its path
as a function of its initial momentum.
The resolution to the particle dE/dx is shown 
in Fig.~\ref{fig:dedx_proton_RNN}, 
where one can see that the energy deposited by a proton as a function of the fitted node position is neater and more refined for the NNs compared to the SIR-PF (with all hits as input), in particular for the Transformer that shows the most accurate Bragg peak.
Automatically, this translates into a more performing particle identification capability, as shown in Tab.~\ref{tab:pid} for different particles such as muons, pions, protons and electrons for a wide range of energies. \\

\begin{table}[htb]
    \centering
    \resizebox{1.0\linewidth}{!}{
    \begin{tabular}{llgggg}
    \hline \hline
    & & \multicolumn{4}{c}{\textbf{Truth}} \\
    %
    \multirow{5}{*}{\textbf{Transformer}} & & \CC $p$ & \CC $\pi^{\pm}$ & \CC $\mu^{\pm}$ & \CC $e^{\pm}$ \\
    & $p$    & 0.907 & 0.057 & 0.071 & 0.020 \\
    & $\pi^{\pm}$      & \CC 0.067 & \CC 0.643 & \CC 0.190 & \CC 0.199 \\
    & $\mu^{\pm}$      & 0.007 & 0.041 & 0.595 & 0.009 \\
    & $e^{\pm}$  & \CC 0.019 & \CC 0.259 & \CC 0.144 & \CC 0.772 \\
    \hline
    \multirow{4}{*}{\textbf{RNN}} & $p$    & 0.896 & 0.080 & 0.089 & 0.027 \\
    & $\pi^{\pm}$      & \CC 0.073 & \CC 0.623 & \CC 0.233 & \CC 0.200 \\
    & $\mu^{\pm}$      & 0.006 & 0.036 & 0.506 & 0.007 \\
    & $e^{\pm}$  & \CC 0.025 & \CC 0.261 & \CC 0.172 & \CC 0.766 \\
    \hline
    \multirow{4}{*}{\textbf{SIR-PF (track hits)}} & $p$    & 0.858 & 0.080 & 0.082 & 0.017 \\
    & $\pi^{\pm}$      & \CC 0.103 & \CC 0.606 & \CC 0.310 & \CC 0.237 \\
    & $\mu^{\pm}$      & 0.014 & 0.042 & 0.453 & 0.006 \\
    & $e^{\pm}$  & \CC 0.025 & \CC 0.272 & \CC 0.155 & \CC 0.740 \\
    \hline
    \multirow{4}{*}{\textbf{SIR-PF (all hits)}} & $p$    & 0.891 & 0.092 & 0.126 & 0.024 \\
    & $\pi^{\pm}$      & \CC 0.077 & \CC 0.603 & \CC 0.236 & \CC 0.229 \\
    & $\mu^{\pm}$      & 0.008 & 0.039 & 0.517 & 0.007 \\
    & $e^{\pm}$  & \CC 0.024 & \CC 0.266 & \CC 0.121 & \CC 0.740 \\
    \hline
    \hline    
   \end{tabular}
   }
    \caption{
    \label{tab:pid} Particle identification (proton $p$, pion $\pi^{\pm}$, muon $\mu^{\pm}$, and electron $e^{\pm}$) confusion matrix for different methods: RNN, Transformer, Sequential Importance Resampling particle filter (SIR-PF) with all hits, and SIR-PF with only track hits as input. Each matrix element corresponds to the probability of correctly identifying an elementary particle.
    Each column of the confusion matrix is normalized to 1 and represents the true particles, whereas the rows represent the predictions.
    }
\end{table}

\section{Discussion}
\label{sec:}



Deep learning is starting to play a more relevant role in 
the design
and exploitation of
particle physics experiments, although it is still in a gestation phase within the high-energy physics community.
If the optimal neural network is optimised, deep learning has the unique capability of building a non-linear multi-dimensional MC-based prior probability function with many degrees of freedom (d.o.f.)
that can efficiently and accurately model all the information acquired in a particle physics experiment
and enhance the performance of the particle track fitting and, consequently, its kinematics reconstruction.
Such a level of detail is, otherwise, nearly impossible to incorporate ``by hand'' in the form of, for example, a covariance matrix to be used in a traditional particle filter.
In this work, we show that a Transformer and a RNN can efficiently learn the details of the particle propagation in matter mixed with the detector response
and lead to a significantly improved reconstruction of the interacting particle kinematics.
We observed that the NNs capture better the details of the particle propagation even when its complexity increases, which is the case near the presence of clusters of hits, for example, due to $\delta$-rays.

It is worth noting that, as mentioned in Sec.~\ref{sec:results}, this work does not aim to report on the performance of the simulated particle detector but rather to show the added value provided by a NN-based fitting.
Moreover, the proposed method does not replace the entire chain of algorithms traditionally adopted in a particle flow analysis (e.g., minimum spanning tree, vertex fitting, etc.) but is meant to assist and complement them as a more performing fitter.
For instance, a possibility could be to apply SIR-PF several times with ``ad-hoc'' manipulation of the data between each step.
However, this would be an unfair comparison as one could also implement multiple deep learning methods and focus on their optimisation.

We believe this approach is a milestone in artificial intelligence applications in HEP and can play the role of a game changer by shifting the paradigm in reconstructing particle interactions in the detectors. 
The prior, which is consciously built from the modelling of the underlying physics from data external to the experiment,
becomes as essential as the real data collected for the physics measurement. De facto, the prior provides a strong constraint to the ``interpretation'' of the data, helping to remove outliers introduced by detector effects such as 
from the smearing introduced by the point spread function
and improving the spatial resolution well below the actual granularity of the detector.

Its accuracy also depends on the quality of the training sample, i.e. on the capability of the MC simulation to correctly reproduce the data. Although this is true for most of the charged particles, a careful characterisation of the detector response will be crucial to validate and, if necessary, tune the simulation (e.g., electromagnetic shower development or hadronic secondary interactions) used to generate the training sample.

This study requires that, first, the signatures observed in the detector are analysed, and the three-dimensional hits that compose tracks belonging to primary particles (directly produced at the primary interaction vertex) are distinguished and analysed independently.
This approach is typical of particle flow analyses.


This work is focused on physics exploitation in particle physics experiments.
However, the developed AI-based methods can also fulfil the requirements in applications outside of HEP, as long as one has a valid training dataset.
One example is proton computed tomography~\cite{https://doi.org/10.1118/1.1884906,pCT-1,pCT-2,PETTERSEN201751} 
used in cancer therapy, where scintillator detectors are used to measure the proton stopping power along its track in the Bragg peak region to precisely predict the stopping position of the proton in the human body. This measurement is analogous to the momentum regression described in Sec.~\ref{sec:reco-kinematics-method}, given the nearly complete correlation between the particle range and momentum.



Future improvements to the developed NNs may involve the direct computation of the node stopping power from the track, i.e., the combined fitting of both the node particle position and energy loss.

\section{Methods}
\label{sec:methods}

\subsection{Description of the fitting algorithms}

\begin{figure*}[htb]
\centering
{\adjincludegraphics[height=10.5cm,trim={0 0 0 0},clip]{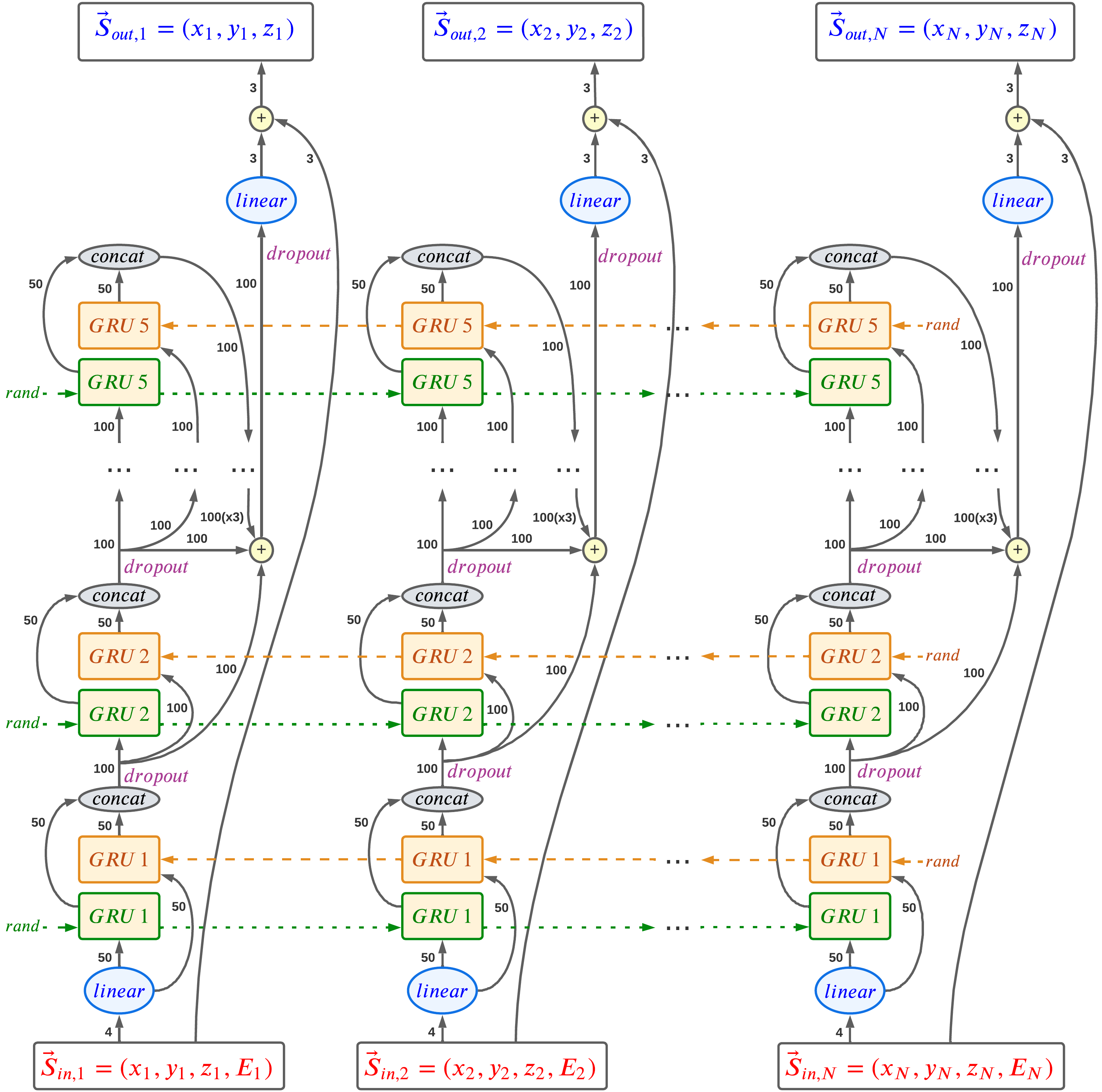}
\hfill
\adjincludegraphics[height=10.5cm,trim={0 0 0 0},clip]{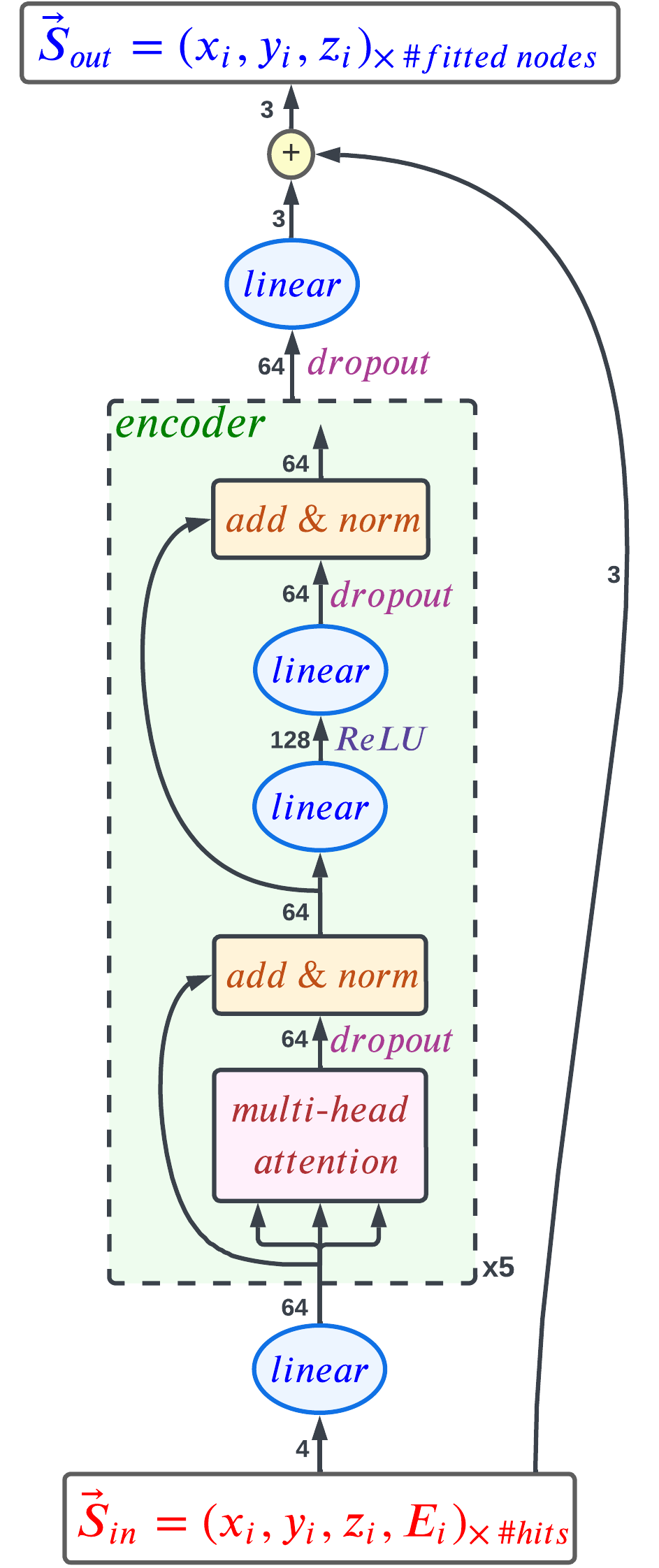}}
  \caption{\label{fig:architectures}
    The architectures of the neural networks implemented: recurrent neural network (RNN, left) and Transformer (right). In high-level terms, RNN consists of five bidirectional GRU layers, while the Transformer consists of five sub-encoder layers. Both models are followed by a linear layer that projects the sum of the outputs of the GRU/encoder layers into a vector of length three. Finally, the input hit position $(x_{i}, y_{i}, z_{i})$ is summed to the network's output, allowing it only to learn the ``residuals'' of the reconstructed hits concerning the true node states ($\vec{S}_{in}\rightarrow\vec{S}_{out}$).}
\end{figure*}

To test the capability of deep learning to fit particle trajectories using reconstructed hits as input, we developed two neural networks that represent the state-of-the-art in the field of natural language processing (NLP, as detailed in the Supplementary Information): the recurrent neural network (RNN)~\cite{JORDAN1997471, 10.5555/553011, SHERSTINSKY2020132306} and the Transformer~\cite{Vaswani2017attention} (see Fig.~\ref{fig:architectures} for a full picture of the architectures). Both algorithms learn from input sequences, each of these sequences being, for instance, a succession of words forming a sentence in the NLP case; or reconstructed hits representing a detected elementary particle in our scenario. Their power rely on their capacity of learning relations between all elements of a sequence. In general terms, RNNs count with memory mechanisms to use information from the ``past'' (previous items in the sequence) and the ``future'' (following items in the sequence) to make predictions. Thus, RNNs assume the input sequences to be ordered. On the other hand, Transformers do not necessarily need sequences to be ordered: the correlations among different items in the sequence are learnt throughout the training process.

We implemented a bi-directional RNN, and the memory mechanism used is the gated recurrent unit (GRU)~\cite{cho-etal-2014-properties}. Our RNN consists of five bi-directional GRU layers with 50 hidden units each. The output of each GRU layer is the concatenation of the forward and backward modules of the layer and is given as input for the following layer (except for the last layer). Instead of propagating only the output of the last GRU layer to the final dense layer, the outputs of all layers are summed together, replicating the concept of ``skipped connections'' in a similar way to what the ResNet or DenseNet model do~\cite{He-et-al-2015-deep}. As regularisation, a dropout of 0.1 is applied to the output of each GRU layer (except for the last GRU layer) and to the summed output of the GRU layers, which is then projected through a final dense layer to have fitted nodes of size 3, representing the coordinates in a three-dimensional space ($x$, $y$, and $z$). The implemented RNN has a total of 213,553 trainable parameters.

The Transformer model designed consists of 5-stacked Transformer-encoder layers, with 8 heads per layer and a dimension of 128 for the hidden dense layer. The input hits are embedded into vectors of size 64. A dropout of 0.1 is applied in each encoder layer and also to the output of the encoder layers to be further projected through a final dense layer (analogously to the RNN), making each fitted node have a length of three. There is no positional encoding since the goal is to make the network learn the relative ordering of the hits based on the 3D positions. The network has a total of 167,875 trainable parameters

\begin{figure*}[htb]
    {\adjincludegraphics[height=4.35cm,trim={{.01\width} 0 {.0\width} 0},clip]{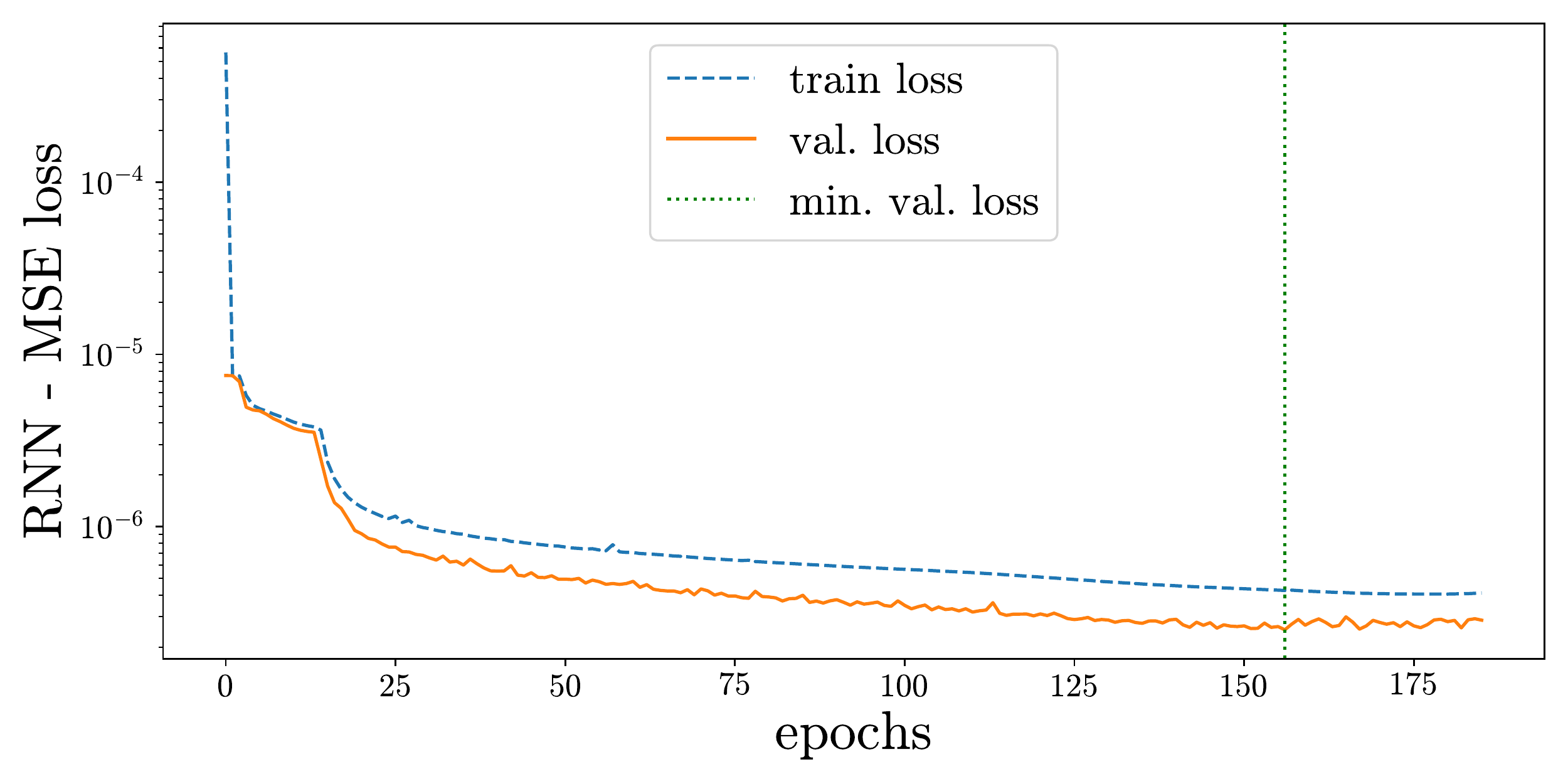}}
    \hfill
    {\adjincludegraphics[height=4.35cm,trim={{.01\width} 0 {.0\width} 0},clip]{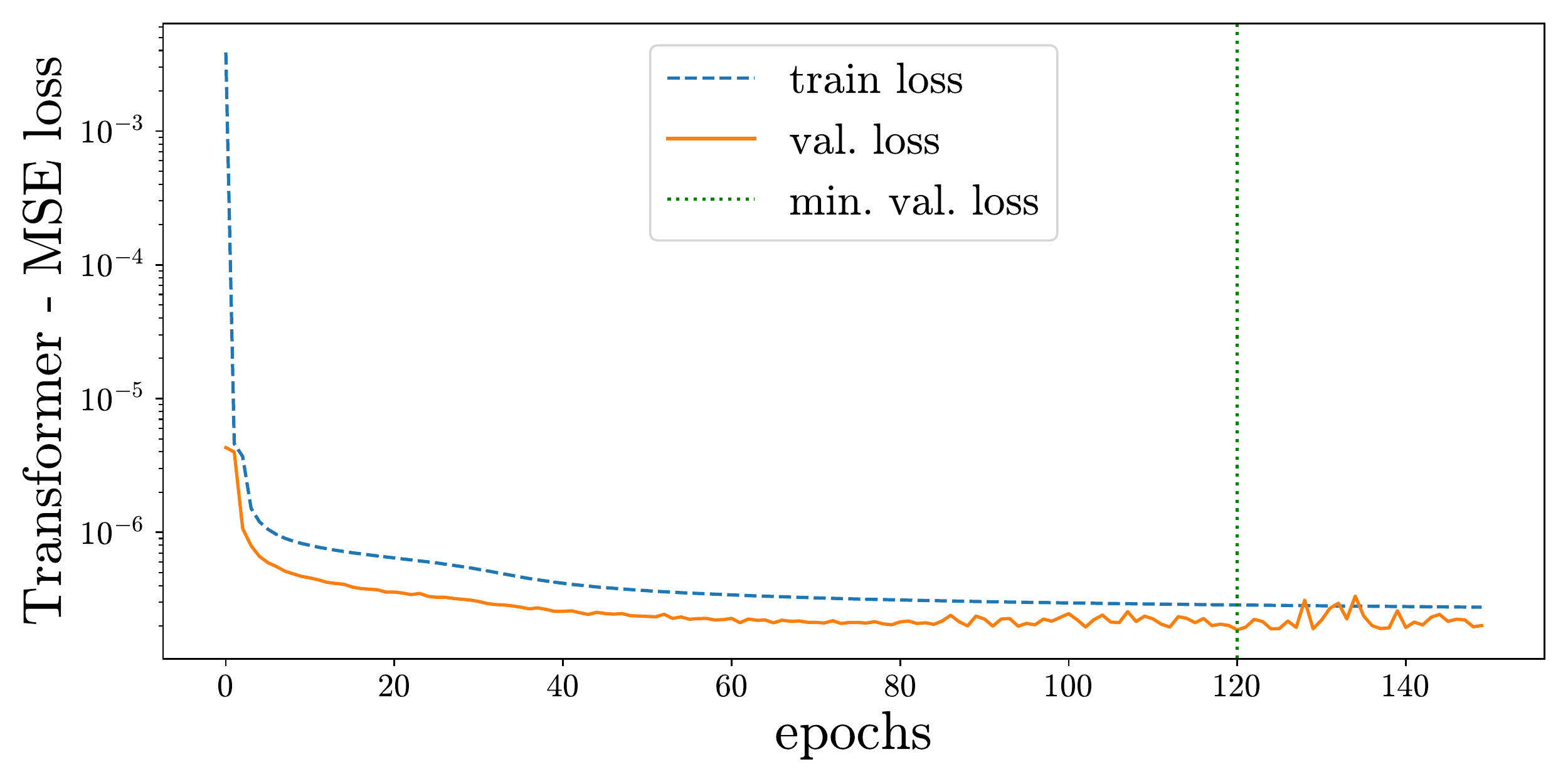}}
\caption{
\label{fig:losses} 
Training and validation loss curves for the recurrent neural network (RNN, left) and the Transformer (right). The loss function used is the mean-squared error (MSE). The dashed-vertical lines represent the epoch that minimises the loss and, thus, the model weights used for the subsequent analysis. The Transformer network converges much faster than the RNN, presumably because the former can learn the correlations among unordered reconstructed hits, and the latter assumes the reconstructed hits are ordered, which can lead to confusion due to the inherent flaws of the ordering provided (impossibility of arranging an optimal order from reconstructed information)}.
\end{figure*}

We implemented both networks in Python v3.10.4~\cite{Python} using PyTorch version 1.11.0~\cite{PyTorch}, and trained them on a dataset of simulated elementary particles consisting of 1,898,620 particles (425,575 protons, 489,978 pions, 492,546 muons and antimuons, and 490,521 electrons and positrons). Each particle consists of a sequence of reconstructed hits with their known positions (centre of the matching cubes) and energy depositions (in an arbitrary signal unit) represented for each hit with the tuple $\vec{S}_{in}=(x_i$, $y_i$, $z_i$, $E_i)$ and truth node position to be learnt $\vec{S}_{out}=(x_i$, $y_i$, $z_i)$. Each variable is normalised to the range [0,1]. We used 80\% of the particles from this sample for training and 20\% for validation, ignoring particles with either less than 10 reconstructed hits or less than 2 track hits, both representing less than 1\% of the total particles. Note that this dataset is statistically independent of the one used for producing the results shown in Sec.~\ref{sec:results}. Mean-squared error and Adam (batch size of 128, learning rate of $10^{-4}$, $\beta_{1}=0.9$, and $\beta_{2}=0.98$) are the loss function (typical for regression) and optimiser, respectively, chosen for both networks. We trained the models on an NVIDIA A100 GPU for an indefinite number of epochs but with an early stopping of 30, meaning that the training terminates when the loss on the validation set does not improve for 30 epochs. The training and validation losses are shown in Fig.~\ref{fig:losses}.

It is necessary to mention that for both the RNN and the Transformer, we sum together (position-wise) the output of the models for each fitted node and the 3D position of the corresponding reconstructed hit given as input. In that way, we force the networks to learn the residuals between reconstructed hits and fitted nodes (in other words, what is learnt is how to adjust each reconstructed hit to a node position that matches the actual particle trajectory).

Regarding the Sequential Importance Resampling particle filter (SIR-PF), for each particle, we use the first reconstructed hit as prior\footnote{Hits are reordered with respect to the axis the particle is travelling through the furthest; if there are several candidates for the first position, we chose the one with the highest energy deposition.}, meaning we use it to sample the first random particles inside that cube, and the energy deposition of each particle happens to be the one of the hitting cube. In each step, the random particles are propagated through the next 15 hits\footnote{We make sure the random particles are sampled inside the available reconstructed hits.} (starting with counting from the position of the current state). For each random particle, the algorithm calculates the variation in $x$, $y$, $z$, $\theta$ (elevation angle defined from the XY-plane, in spherical coordinates), and energy deposition (in an arbitrary signal unit) between the particle and the current state and assigns a likelihood based on the value of the selected bin in a 5-dimensional histogram\footnote{The histogram, used for the likelihood calculation of the SIR-PF, is filled with the variation between consecutive true nodes in $x$, $y$, $z$, $\theta$, and energy deposition, named: $\Delta x$, $\Delta y$, $\Delta z$, $\Delta \theta$, and $\Delta E$, respectively. The histogram has 100 bins per dimension.}, pre-filled using the same dataset used to train the RNN and the Transformer. In that way, the next state ends up being the weighted average (using the pre-computed likelihood) of the positions of the different sampled particles available. The filter is run from the start to the end of the particle (forward fitting) and from the end to the start (backward fitting); the results of the forward and backward fittings are averaged in a weighted manner, giving more relevance to nodes fitted last in both cases. The total number of random particles sampled in each step is 10,000.

\subsection{Computation of particle kinematics}
\label{sec:reco-kinematics-method}


The RNN, Transformer, and SIR-PF outputs are analysed to extract the kinematics from the fitted tracks. The performance of the methods depends on the accuracy of the fitted nodes compared to the true track trajectories. The same procedure has been applied to the nodes fitted with the different algorithms for a fair comparison.

The following steps have been followed to perform the physics analysis, that is, particle identification (PID), momentum reconstruction and charge identification (charge ID):
\begin{enumerate}
    \item 
    Extract ``track'' nodes:
    the input 3D hits can be divided into two categories: 
    (1) track hits, 
    directly crossed by the   
    charged particle,
    (2) crosstalk hits, 
    caused by the leakage of scintillation light from the cube containing the charged particle.
    After the track is fitted, the 3D hits are identified as track-like if there is a scintillator cube with a particular energy deposition that contains the fitted node.
    The remaining nodes are classified as non-track, and they include crosstalk hits. 
    The scintillation light observed in a non-track hit is summed to the nearest track hit. The position of the fitted node is then used to compute the stopping power ($\mathrm{d}E/\mathrm{d}x$).
    
    \item 
    Node energy smoothing: 
    the energy of the remaining ``track'' nodes is smoothed in order to eliminate fluctuations due, for example, to the different path lengths travelled by the particle in the adjacent cubes (the scintillation light in a cube is nearly proportional to the distance travelled by the particle).
    The smoothing of an energy node is performed by applying an average over the energy of nearby nodes weighted by a Gaussian distribution function of the respective distance.
    
    \item 
    Particle identification and momentum regression:
    a gradient-boosted decision tree (GBDT)~\cite{hastie2009boosting}, available in the TMVA package of the CERN ROOT analysis software (\url{https://root.cern.ch/}), was used to perform the particle identification and the momentum regression.
    The GBDT input parameters were chosen as: (1) the first 5 and the last 10 fitted node energies along the track; (2) the neighbouring node distances of those 15 nodes;
    (3) 
    the track total length and energy deposition.
    Two independent GBDTs with the same structure were trained to reconstruct the primary particle type (muon, proton, pion, or electron, classification) and its initial momentum (regression).

\end{enumerate}

The electric charge of the particle was identified by measuring the deflection of the track projected to the plane perpendicular to the magnetic field. 
The convex or concave deflection implies either a positive or a negative charge, where the positions of the fitted nodes were used.





The momentum reconstruction from the track curvature produced by the magnetic field was estimated for the resolutions provided by different track fitters and studied for different configurations by using parameterised formulas that incorporate the spatial resolution from tracking in a magnetic field as well as the multiple scattering in dense material \cite{Gluckstern:1963ng,SperdutiFirstGNN},
that have been shown to reproduce data well enough for sensitivity studies.

\section{Acknowledgements}

Part of this work was supported by the SNF grant PCEFP2\_203261, Switzerland.









\bibliographystyle{ieeetr}
\bibliography{biblio}

\section{Supplementary Information}
\label{sec:supplementary_info}

\subsection*{Natural language processing and deep learning}

Fitting reconstructed hits into a number of nodes that form an approximation to the true track trajectory can be modelled similarly to problems from the field of natural language processing (NLP). In NLP, it is common to work with sequences of words, forming sentences, and the aim is to perform tasks such as text translation, text synthesis, or speech recognition, which require algorithms that have the potential to deal with the possible different relations of entities within a sentence~\cite{James2003NLP, Chowdhary2020, Hirschberg2015}. Analogously, in the problem described in this article, the reconstructed hits can be seen as an ordered (sorted along with one axis) sequence of points, which would make it straightforward for an algorithm brought from NLP to exploit those points and predict the trajectory of the track through the detector.

Nowadays, artificial intelligence (AI) is the leading choice for handling the vast majority of NLP problems, offering sophisticated algorithms that have set unprecedented results in the discipline~\cite{mishra2020natural,LAURIOLA2022443}. Most of these AI algorithms are categorised in the sub-field of deep learning and, more concretely, the family of ``recurrent'' neural networks (RNNs)~\cite{JORDAN1997471, 10.5555/553011, SHERSTINSKY2020132306} stand out. Standard feed-forward neural networks were the initial inspiration for RNNs, but RNNs highlight an extraordinary ability to learn from the semantics of temporal sequences by being trained on large amounts of data.

\subsubsection{Recurrent neural networks}

In contrast to other neural networs, RNNs can handle input sequences of different lengths and share features learnt across different positions within the sequences, mainly thanks to their capacity to use their internal states as ``memory''. Considering an input sequence where each position corresponds to a different time step, a standard RNN unit will produce the following activation $a^{<t>}$ and output $y^{<t>}$ for the input position $x^{t}$ of the sequence at time step $t$:

\vspace{-0.1cm}

\begin{equation}
\begin{aligned}
a^{<t>} &= g(a^{<t-1>}, x^{t}; \theta_{a})\\&= g(W_{aa}a^{<t-1>}+W_{ax}x^{t}+b_{a}) 
\end{aligned}
\label{eq:activation}
\end{equation}

\begin{figure}[htb]
\centering
\raisebox{-0.5\height}{\includegraphics[width=0.65\linewidth]{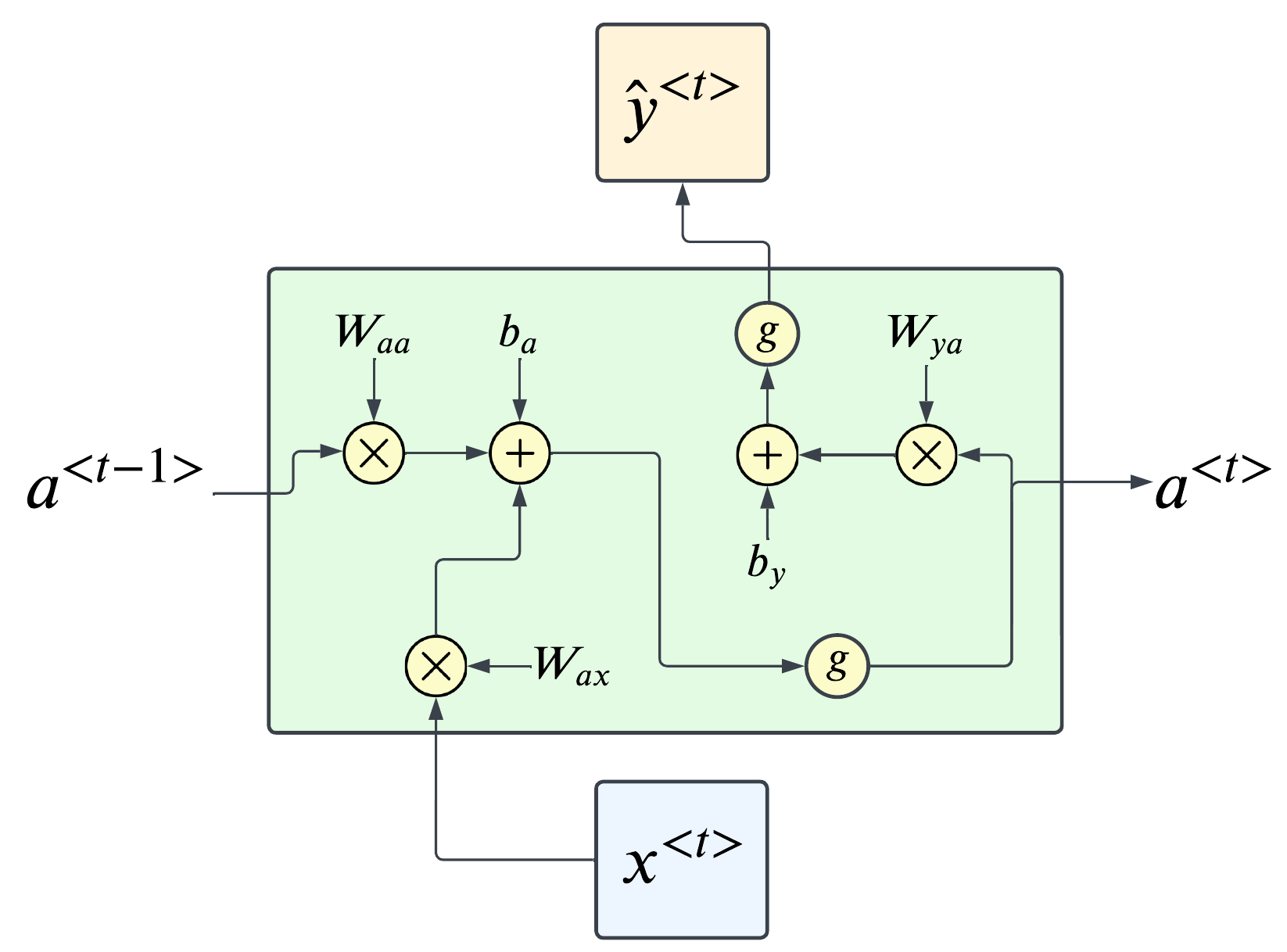}}\\
\vspace{0.5cm}
\raisebox{-0.5\height}{\includegraphics[width=0.8\linewidth]{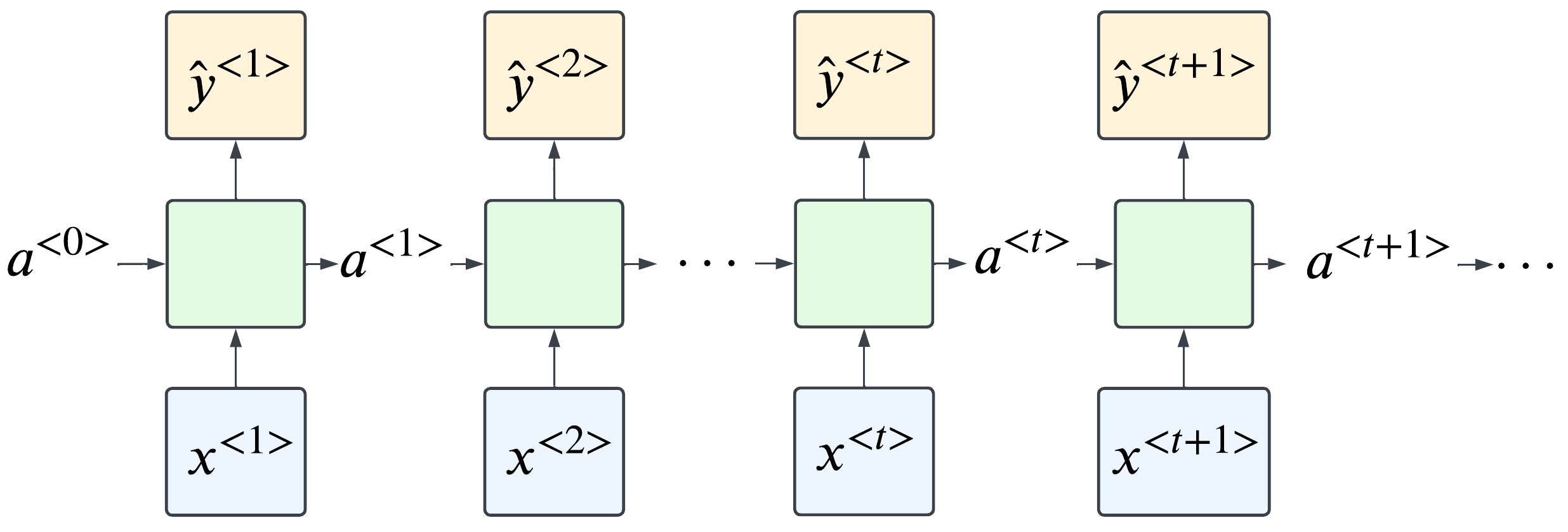}}
\caption{\label{fig:rnn_structure} (Top) Internal structure of a recurrent neural network (RNN) unit for the time step $t$, where the input position $x^{t}$ of the sequence and the previous activation $a^{<t-1>}$ are used to calculate the next activation $a^{<t>}$ and output $\hat{y}^{<t>}$; (bottom) unfolded structure of a standard RNN, where the activation at one time step becomes an input to the next time step.}
\end{figure}

\vspace{-0.35cm}

\begin{equation}
\hat{y}^{<t>} = g(a^{<t>}; \theta_{\hat{y}}) = g(W_{\hat{y}a}a^{<t>}+b_{\hat{y}}) 
\label{eq:output}
\end{equation}

where $g$ is the activation function (e.g, hyperbolic tangent or ReLU), $a^{<t-1>}$ is the activation at time step $t-1$, and $\theta_{a}$ and $\theta_{\hat{y}}$ are the network parameters needed for calculating $a^{<t>}$ (i.e., $W_{aa}$, $W_{ax}$, and $b_{a}$) and $\hat{y}^{<t>}$ (i.e., $W_{\hat{y}a}$ and $b_{\hat{y}}$), respectively. Note that, for each time step $t$, the network is not only using the position $x^{t}$ of the sequence as input but also the activation of the immediate previous time step to calculate the next activation and output. In this way, RNNs can reuse previous activations to learn about temporal information. This behaviour is depicted graphically in Fig.~\ref{fig:rnn_structure}. It is relevant to mention that the network parameters are shared over time, meaning that the model size does not increase with the length of the input sequence.

Having the output of the network $\hat{y}$ and the true labels $y$, the discrepancy between the two is evaluated with the following loss function $\mathcal{L}$:

\begin{equation}
L(\hat{y}, y) = \frac{1}{T} \sum_{t=1}^{T} \mathcal{L}(\hat{y}, y)
\label{eq:loss}
\end{equation}

where T is the total number of time steps. In RNNs, the model weights $\theta$ are updated during backward propagation at each time step, what is generally called backpropagation through time:

\begin{equation}
\frac{\partial \mathcal{L}}{\partial \theta} = \frac{1}{T} \sum_{t=1}^{T} \frac{\partial \mathcal{L}(\hat{y}_{t}, y_{t})}{\partial \theta}
\label{eq:backprop}
\end{equation}

\begin{figure}[htb]
\centering
{\includegraphics[width=0.9\linewidth]{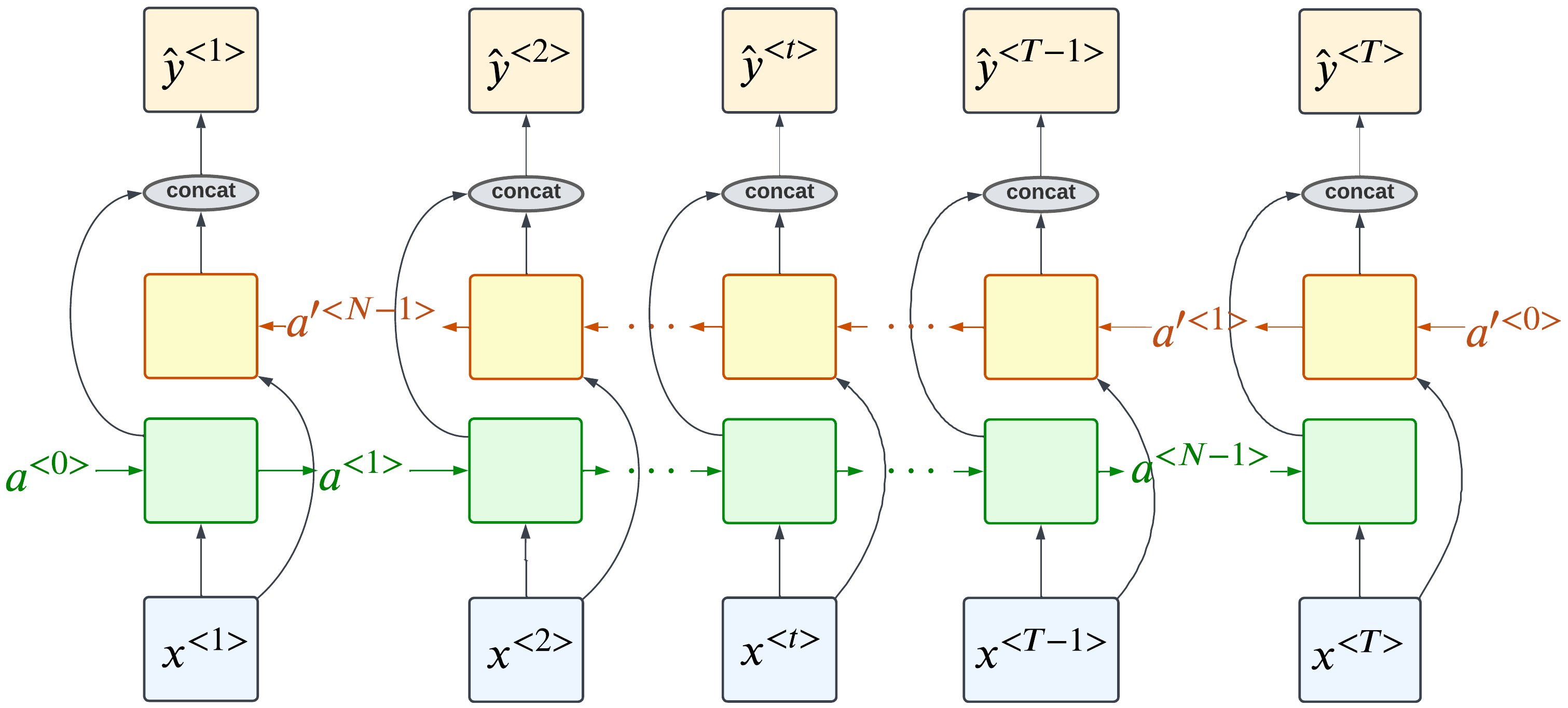}}
\caption{\label{fig:brnn} Overview of a standard bidirectional recurrent neural network (BRNN) architecture. The architecture consists of two RNNs combined together. The input sequence is given from start to end (from left to right in the figure, in green) to one of them, and from back to front (right to left, in yellow) to the other one. The output of each RNN is normally concatenated for each time step.}
\end{figure}

In the above scenario, in order to make predictions on the current position, the model can learn about the previous part of the sequence. However, all the following positions are ignored. In other words, the model has the ability to learn from the ``past'' but not from the ``future''. Accessing future information might be necessary to report accurate results in some cases. For example, coming back to the physics problem presented in this manuscript, to precisely predict the closest 3D position to the actual particle trajectory for a particular reconstructed hit, it might be advantageous to access both the previous and the next hits within the sequence. The solution to also learn about the future is to put two independent RNNs together into what is called a bidirectional recurrent neural network (BRNN)~\cite{Schuster1997bidirectional}, where the input is given from start to end to one RNN and from back to the front to the other RNN; then, the outputs at each time step usually are concatenated, as illustrated in Fig.~\ref{fig:brnn}. In this way, for each time step, the network has access to the activations coming from the previous position and the following position in the sequence, giving the model the ability to learn from the past and the future simultaneously. 

Figures~\ref{fig:rnn_structure} and~\ref{fig:brnn} show the case where the length of the input sequence matches the length of the output; one example of this could be a problem where the goal is to categorise each word in a sentence into the corresponding category (e.g., noun, pronoun, verb, or adjective). Another example, which is solved in this manuscript, is to predict the closest track trajectory point for each input reconstructed hit. Nevertheless, there are many other RNN topologies: many-to-one, where only the output of the last time step is considered (e.g., for sentiment classification); or many-to-many, but, in this case, the length of the output sequence does not necessarily have to match the length of the input sequence (e.g., text translation or music generation).

\subsubsection{GRU and LSTM}

Some sequence models might be affected by very long-term dependencies, meaning that, within a sequence, it could be possible to find strong relations such as the dependency of an arbitrary position $i$ and a position $i+k$, being $k$ a large positive integer. On top of that, since the input sequences can have different lengths, the long-term dependencies can be arbitrarily long. 

Due to the continuous recalculation of the activations (shown in Equation~\ref{eq:activation}), standard RNNs are not good at catching long-term dependencies, arising vanishing/exploding gradient problems during back-propagation~\cite{Bengio1994learning,hochreiter2001gradient,pascanu2012diff}. Several architectures have been proposed to deal with this issue, where gated recurrent and long short-term memory units stand out.

A gated recurrent unit (GRU)~\cite{Cho2014learning, Dey2017GRU,cho-etal-2014-properties,Chung2014gru} is an alternative to the original RNN approach that handles long-term dependencies by calculating a candidate $\tilde{a}^{<t>}$ of the activation (Eq.~\ref{eq:gru_candidate}) using a gate to measure how relevant the previous activation is to compute the next candidate (Eq.~\ref{eq:gru_gate_r}).

\begin{equation}
\begin{aligned}
\tilde{a}^{<t>} &= tanh(a^{<t-1>}, x^{t}; \theta_{a})\\ &= tanh(W_{aa}(\Gamma_{r}\odot a^{<t-1>})+W_{ax}x^{t}+b_{a})
\end{aligned}
\label{eq:gru_candidate}
\end{equation}

\begin{figure}[htb]
\centering
{\includegraphics[width=0.9\linewidth]{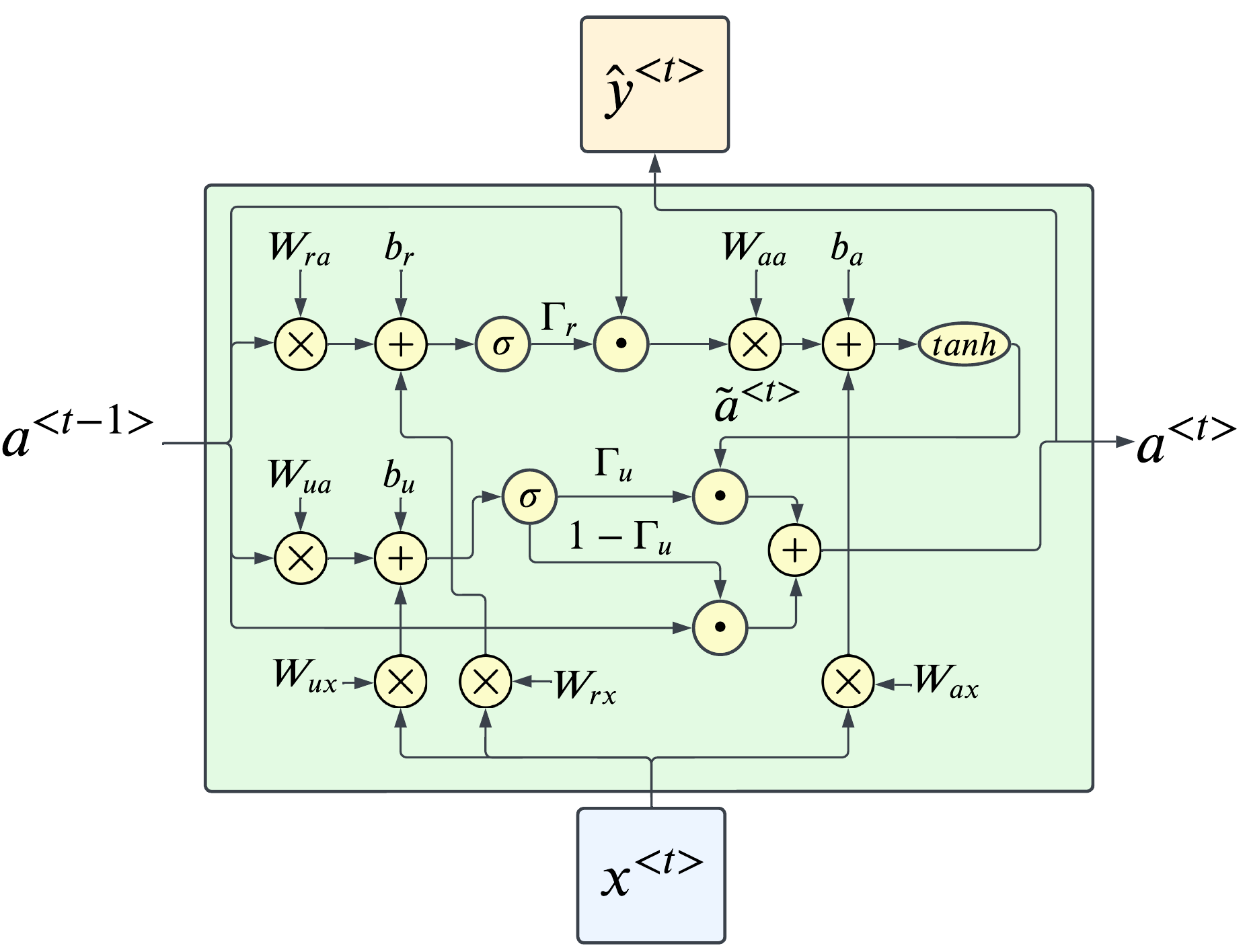}}
\caption{\label{fig:gru} Gated recurrent unit (GRU) structure. For each time step, the input $x^{<t>}$ and the previous activation $a^{<t-1>}$ are used to compute the new activation $a^{<t>}$ and the output $\hat{y}^{<t>}$. To do so, the gate $\Gamma_{r}$ is used to give relevance to the previous activation when calculating the activation candidate $\tilde{a}^{<t>}$, while the gate $\Gamma_{u}$ is used to update the activation by weighting the candidate $\tilde{a}^{<t>}$ and the previous activation $a^{<t-1>}$.}
\end{figure}

\vspace{-0.35cm}

\begin{equation}
\Gamma_{r} = \sigma(W_{ra}a^{<t-1>}+W_{rx}x^{t}+b_{r})\\
\label{eq:gru_gate_r}
\end{equation}

where $tanh$ and $sigma$ are the hyperbolic tangent function and the sigmoid function, respectively. The activation is then updated using another gate (Eq.~\ref{eq:gru_gate_u}) to weight the candidate and the previous activation into the new activation (Eq.~\ref{eq:update}). Figure~\ref{fig:gru} represents the GRU workflow as a whole.

\begin{equation}
\Gamma_{u} = \sigma(W_{ua}a^{<t-1>}+W_{ux}x^{t}+b_{u})\\
\label{eq:gru_gate_u}
\end{equation}

\vspace{-0.35cm}

\begin{equation}
a^{<t>} = \Gamma_{u}\odot\tilde{a}^{<t>} + (1-\Gamma_{u})\odot a^{<t-1>}
\label{eq:update}
\end{equation}

\begin{figure}[ht]
\centering
{\includegraphics[width=0.9\linewidth]{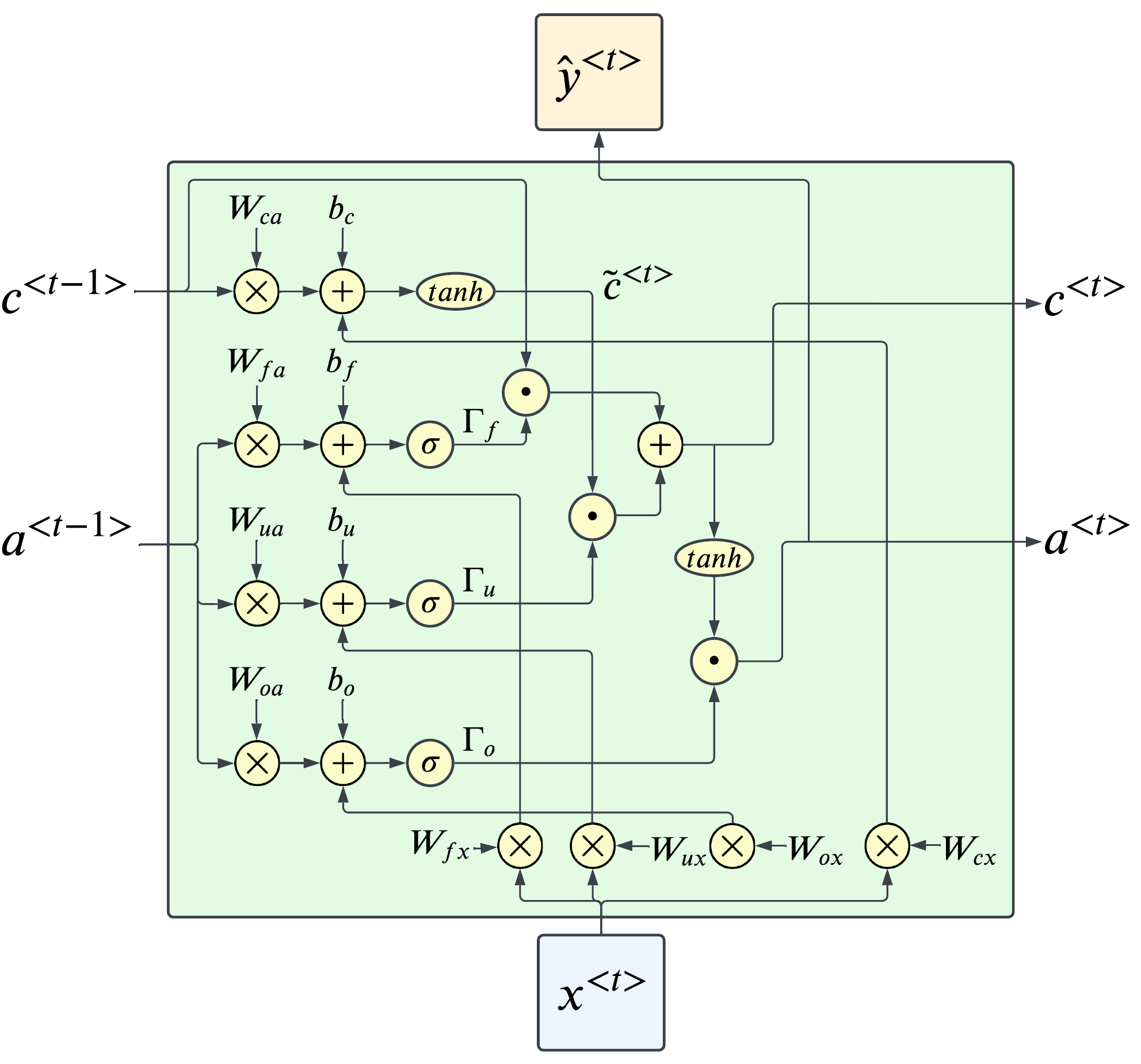}}
\caption{\label{fig:lstm} Long short-term memory (LSTM) unit. For each time step, the input $x^{<t>}$, and the previous activation $a^{<t-1>}$ and memory cell $c^{<t-1>}$ are used to compute the new activation $a^{<t>}$, memory cell $c^{<t>}$, and output $\hat{y}^{<t>}$. The gates $\Gamma_{u}$ and $\Gamma_{f}$ contribute to the computation of the new memory cell $c^{<t>}$, while the gate $\Gamma_{o}$ is used to calculate the activation $a^{<t>}$.}
\end{figure}

Similarly to GRU, the long short-term memory (LSTM)~\cite{LSTM, Greff2017lstm, Yu2019review} unit handles long-term dependencies by not only updating the activation $a^{<t>}$ at each time step but also updating a new entity named the ``memory'' cell $c^{<t>}$. The LSTM unit uses three different gates: (1) a gate $\Gamma_{u}$ (Eq.~\ref{eq:gru_gate_u}, equivalent to th GRU version) that tells how much the memory cell candidate $\tilde{c}^{<t>}$ (Eq.~\ref{eq:lstm_candidate}) should affect the update of the new memory cell $c^{<t>}$ (Eq.~\ref{eq:lstm_update_c}); (2) a gate $\Gamma_{f}$ (Eq.~\ref{eq:lstm_gate_f}) that measures how much to forget about the previous memory cell $c^{<t-1>}$ during the new memory cell $c^{<t>}$ calculation; and (3) a gate $\Gamma_{o}$ (Eq.~\ref{eq:lstm_gate_o}) that weights the memory cell during the calculation of the new activation $a^{<t>}$ (Eq.~\ref{eq:lstm_update_a}). Figure~\ref{fig:lstm} helps understand the above formulas for the LSTM unit by showing the different calculations in a diagram.

\begin{equation}
\begin{aligned}
\tilde{c}^{<t>} &= tanh(c^{<t-1>}, x^{t}; \theta_{c})\\ &= tanh(W_{cc}(\Gamma_{r}\odot c^{<t-1>})+W_{cx}x^{t}+b_{c})
\end{aligned}
\label{eq:lstm_candidate}
\end{equation}

\vspace{-0.35cm}

\begin{equation}
c^{<t>} = \Gamma_{u}\odot\tilde{c}^{<t>} + \Gamma_{f}\odot c^{<t-1>}
\label{eq:lstm_update_c}
\end{equation}

\vspace{-0.35cm}

\begin{equation}
\Gamma_{f} = \sigma(W_{fa}a^{<t-1>}+W_{fx}x^{t}+b_{f})\\
\label{eq:lstm_gate_f}
\end{equation}

\vspace{-0.35cm}

\begin{equation}
\Gamma_{o} = \sigma(W_{oa}a^{<t-1>}+W_{ox}x^{t}+b_{o})\\
\label{eq:lstm_gate_o}
\end{equation}

\vspace{-0.35cm}

\begin{equation}
a^{<t>} = \Gamma_{o}\odot tanh(c^{<t>})
\label{eq:lstm_update_a}
\end{equation}

In practice, both GRU and LSTM perform similarly in terms of the quality of the results for different problems~\cite{Fu2016using,Yin2017comp,Shewalkar2019,Yamak2019comp}. However, due to the lack of a memory unit and thus requiring fewer calculations, GRU tends to be the preferred choice over LSTM since the former is more computationally efficient~\cite{gruvslstm,Yang2020lstmgru}.

\subsubsection{Transformers}

Even though RNNs and their variants GRU and LSTM have reported remarkable results in the field of NLP, they still face a couple of drawbacks. On the one hand, sequences are processed position by position and not altogether, with the risk of still forgetting information regardless of the memory mechanisms, which might not retain all the necessary relations among positions within the sequence. On the other hand, in order to learn about the ``past'' and the ``future'' for each time step, bidirectional models are needed, which require twice the usual computation.

A Transformer~\cite{Vaswani2017attention} is a type of neural network, initially proposed for text translation but with many different current applications, that resolves the issues above by treating each input sequence as a whole. Its main feature is the multi-head self-attention mechanism (revolutionising the attention proposed in~\cite{Bahdanau2014neural} and~\cite{Luong2015Effective}), which decides the fragments of the input sequence that are more relevant for the target task by capturing correlations among all items in a sequence. Formally, an input sequence $X\in R^{N \times d_k}$ (sequence of length $N$, each position represented by $d_k$ values) is multiplied by three weight matrices $W_Q$, $W_K$, and $W_V \in R^{d_k \times d_{model}}$ (where $d_{model}$ the length of the new representation for each position of the sequence after the attention mechanism) to produce $Q$ (queries), $K$ (keys), and $V$ (values) $\in R^{N \times d_{model}}$, respectively. The self-attention is calculated with the following formula:

\begin{equation}
Attention(Q,K,V) = softmax(\frac{QK^T}{\sqrt{d_k}})V
\label{eq:attention}
\end{equation}

\begin{figure}[htb]
\centering
{\includegraphics[width=0.65\linewidth]{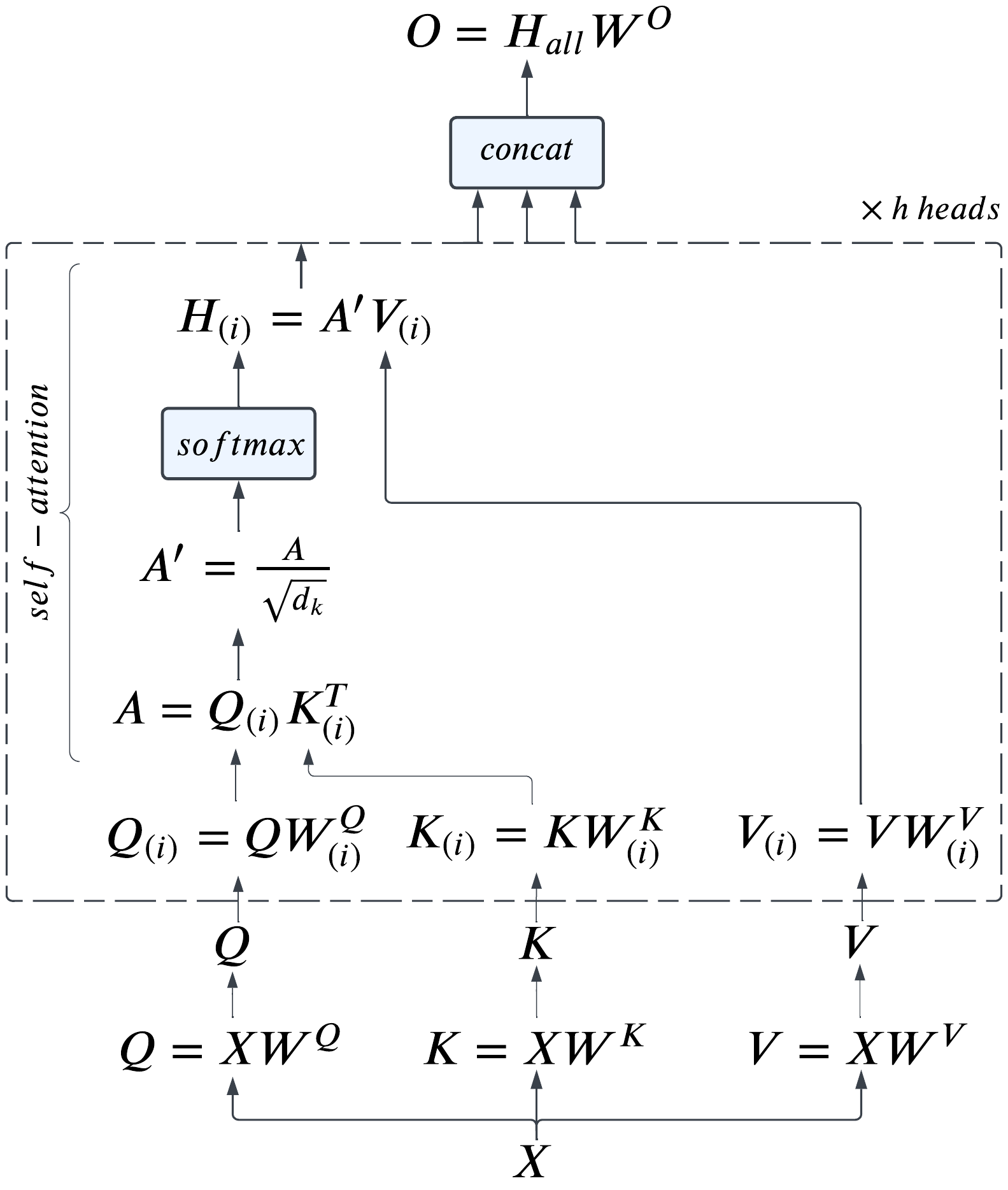}}
\caption{\label{fig:attention} Multi-head attention. The input $X$ is linearly projected into $Q$, $K$, and $V$ (the initial $W^{Q}$, $W^{K}$, and $W^{V}$ matrices are usually the same, resulting in equivalent $Q$, $K$, and $W$; in NLP problems, this first linear projection is called ``embedding''), which are processed $h$ times (one per head) through a self-attention mechanism. The output of the heads are concatenated and multiplied by a final weight matrix $W^O$ to produce the output O. }
\end{figure}

The multi-head part implies repeating the self-attention $h$ times (each self-attention calculation with independent learnt parameters is known as a ``head'') on $Q$, $K$, and $V$ projected through $h$ sets of weight matrices $W^Q_{(i)}$, $W^K_{(i)}$, and $W^V_{(i)} \in R^{d_{model}\times d_k}$ (in the multi-head approach, $d_k = d_{model}/h$.). Then, the outputs of the heads are concatenated and multiplied by a final weigh matrix $W^O \in R^{hd_{k}\times d_{model}}$, as shown in Eqs.~\ref{eq:multihead} and \ref{eq:multihead_attention} and Fig.~\ref{fig:attention}:

\begin{equation}
Multi\mbox{-}head(Q, K, V) = concat(head_{(1)}, ..., head_{(h)})W^O
\label{eq:multihead}
\end{equation}

where:

\begin{equation}
head_{(i)} = Attention(QW^Q_{(i)}, KW^K_{(i)}, VW^V_{(i)})
\label{eq:multihead_attention}
\end{equation}

In order to preserve the order of the input sequence through the different projections and let the model learn about relative positions, ``positional encodings'' are summed to the first linear projection of the input. The authors of the Transformer model chose sine and cosine functions of different frequencies for the positional encoding~\cite{Vaswani2017attention}:

\begin{equation}
\label{eq:neuron3}
PE_{(pos, i)}=\begin{cases}sin(\frac{pos}{10000^{i/d_{model}}}) & \textrm{if}\ i\ \textrm{is even} \\
cos(\frac{pos}{10000^{(i-1)/d_{model}}}) & \textrm{if}\ i\ \textrm{is odd} \\\end{cases}
\end{equation}

where $pos$ is the position in the sequence and $i$ is the dimension.

\begin{figure}[h]
\centering
{\includegraphics[width=0.6\linewidth]{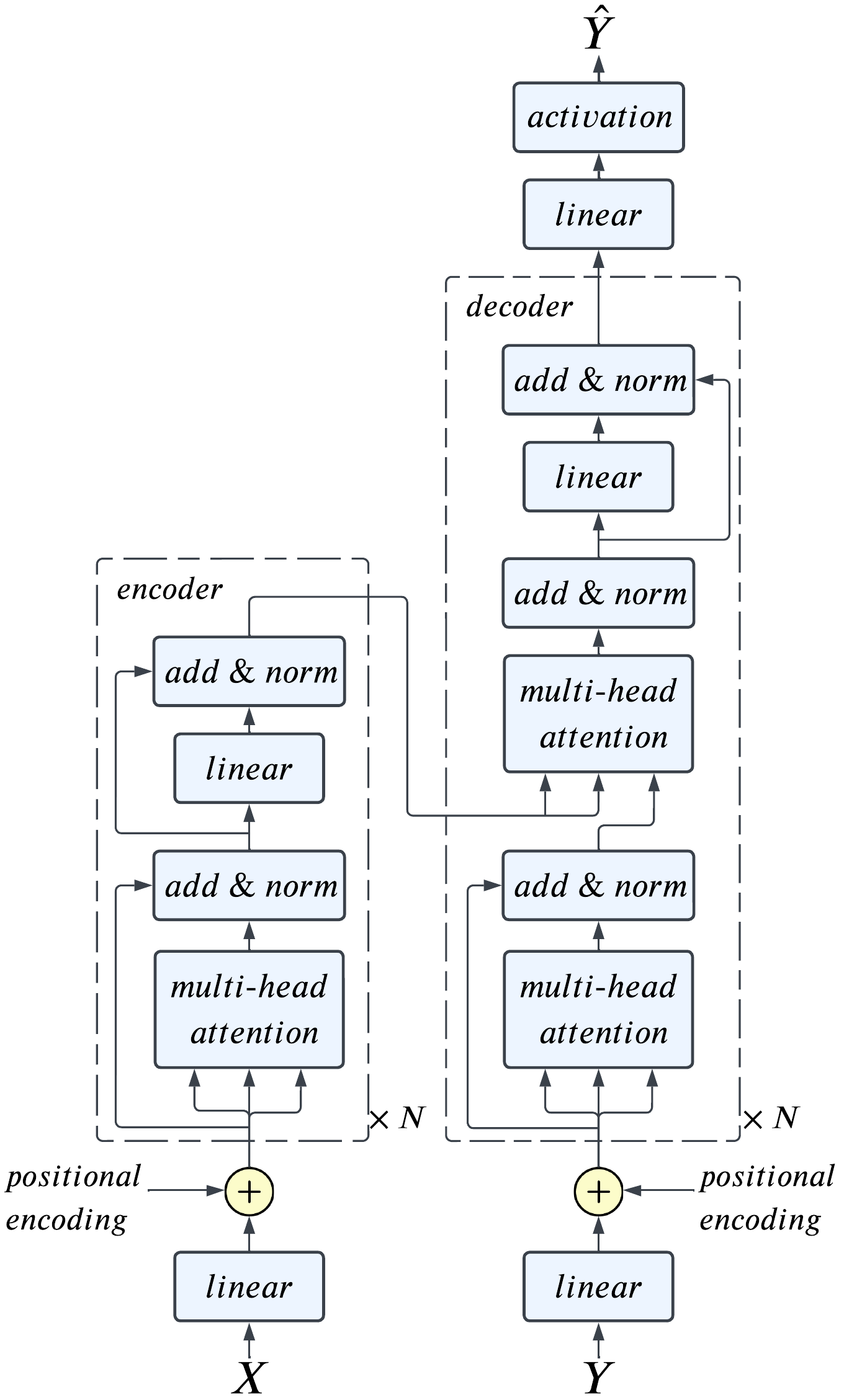}}
\caption{\label{fig:transformer} Transformer encoder-decoder architecture. Figure based on the model published in~\cite{Vaswani2017attention}.}
\end{figure}

Before putting everything together, it is advised to mention that the original Transformer architecture consists of two main components: an encoder and a decoder. In machine translation, the encoder learns relevant features from the input sequence that are useful for the decoder to generate the translated sequence sequentially. In the Transformer model, the inputs are first projected through a linear layer in addition to applying the positional encoding to finally go through the encoder, which consists of a multi-head attention module, an addition (of the output and the input of the multi-head attention) and a normalisation, followed by another linear layer and a final addition+normalisation, all repeated N times. Similarly, the outputs (shifted right) are projected through a linear layer. A positional encoding is applied, to then go through the decoder, which consists of a multi-head attention module, an addition+normalisation, another multi-head attention module (where the input queries $Q$ and keys $K$ are the output of the encoder, which lets the decoder decide which encoder input is relevant for the decoding task), another addition+normalisation, a linear layer, and a final addition+normalisation, all repeated N times. The procedure described is depicted in Fig.~\ref{fig:transformer}.

The original architecture needs the decoder part since it was designed for machine translation. However, for other problems, such as sentiment analysis, the decoder can be replaced with a simpler module (e.g., a linear layer)~\cite{Tan2019Computation,tang-etal-2020-dependency,Devlin2018comp} since there is no need to predict an output sequence but, for example, a single label. In other cases, like the model proposed in this article, the goal is to predict an output for each item in the input sequence; thus, the decoder can be omitted and substituted by a linear layer.




\end{document}